\UseRawInputEncoding
\documentclass[preprint,aps,12pt,nofootinbib,tightenlines]{revtex4}
\usepackage{mathrsfs}
\usepackage{amsmath}
\usepackage{amssymb}
\usepackage{epsfig}
\usepackage{graphicx}
\usepackage{subfigure}
\usepackage{multirow}
\usepackage{booktabs}
\def\be{\begin{eqnarray}}
\def\en{\end{eqnarray}}
\def\non{\nonumber\\}

\usepackage[
 colorlinks=true,
 linkcolor=blue,
 citecolor=blue,
]{hyperref}

\allowdisplaybreaks

\begin{document}
\title{\Large \bf \boldmath $B\to K\bar K(\pi\eta)h$ decays in the presence of isovector scalar resonances $a_0(980,1450)$}

\author{Si-Yang Wang$^1$, 
	    Zhi-Qing Zhang$^2$\footnote{zhangzhiqing@haut.edu.cn}, 
           Zhi-Jie Sun$^3$,
	    Jian Chai$^2$, 
	    Peng Li$^4$\footnote{lipeng@haut.edu.cn} } 

\affiliation{\it \small $^1$ Institute of Particle Physics and Key Laboratory of Quark and Lepton Physics (MOE), Central China Normal University, Wuhan, Hubei 430079, China\\
\it \small $^2$ Institute of Theoretical Physics, School of Sciences, Henan University of Technology, Zhengzhou, Henan 450001, China \\
             \it \small $^3$ Bingtuan Xingxin Vocational and Technical College, Xinjiang, 841007, China\\
             \it \small $^4$ Institute for Complexity Science, Henan University of Technology, Zhengzhou, Henan 450001, China} 


\begin{abstract} \noindent
Different from the previous treatment in a two-body framework, we introduce the dimeson distribution amplitudes (DAs) to describe the strong dynamics between the S-wave resonances $a_0(980, 1450)$ and the $K\bar K (\pi\eta)$ pair, where the Gegenbauer coefficient required is determined from the experimental data on the time-like form factors involved. The branching ratios and direct CP asymmetries of the decays $B \to a^{(\prime)}_0 h \to K\bar K(\pi\eta) h$, with $a_0=a_0(980)$, $a^{\prime}_0=a_0(1450)$ and $h$ referring to a pion or a kaon, are then calculated in the perturbative QCD (PQCD) approach. We find that the branching ratios of the corresponding quasi-two-body decays $B\to a^{(\prime)}_0 K$ obtained with the narrow width approximation are closer to those predicted in the QCD factorization (QCDF) approach compared to the previous PQCD calculations, no matter a three-body or a two-body framework is assumed. Furthermore, all our predictions for these $B\to a^{(\prime)}_0 K$ decays are below the current experimental upper limits except for those of decays $B^0\to a^{(\prime)-}_0K^+$, which are (slightly) larger than the upper limits. Under the narrow width approximation, the branching ratios of the decays $B^+\to a^{(\prime)+}_0\pi^0$, $B^0\to a^{(\prime)+}_0\pi^-$ and $B^0\to a^{(\prime)0}_0\pi^0$ are comparable to or agree well with the previous PQCD and the QCDF calculations. While for the decays $B^+\to a^{(\prime)0}_0\pi^+$ and $B^0\to a^{(\prime)-}_0\pi^+$, their branching ratios are predicted to be unexpectedly large, for example, the obtained branching ratio of decay $B^+\to a^0_0\pi^+$ is even higher than the current experimental upper limit. For the decays $B^0\to a^{(\prime)-}_0K^+, a^{(\prime)0}_0\pi^0, a^{(\prime)+}_0\pi^-$, where the tree and the penguin amplitudes are comparable in magnitude, the direct CP asymmetries are sensitive to the strong phases, which are very different obtained by other theoretical predictions for some channels. Therefore, it is very important to accurately determine the strong phases of these decays, which are crucial to precisely predict the direct CP asymmetries and further clarify the inner structures of the resonances $a_0(980)$ and $a_0(1450)$.   
\end{abstract}

\maketitle

\section{Introduction}
\label{intro}

Although many scalar mesons have been well established in experiments, their spectrum and structure are always an interesting topic from both the experimental and theoretical aspects. In particular, the nature of the light scalar mesons still 
remains a puzzle for the nonperturbative characteristics of QCD at low energy. Among them, the isovector $a_0(980)$ and the isoscalar $f_0(980)$ are most striking~\cite{Hanhart,Sekihara,wwang}. Various phenomena show that these two scalar mesons with degenerate masses can mix with each other through isospin-violating effect~\cite{close,achasov1,wujj}, and both of them can couple to the $K \bar K$ state. Up to now, these light scalar mesons have been described as the quark-antiquark pairs, the tetraquarks, the $K\bar K$ molecules, the hybrid states and so on. In the
$q\bar q$ picture, the isovector $a_0(980)$ is viewed as a P-wave state, whose flavor wave functions are given as
\be
\left|a^0_0(980)\right\rangle=\frac{1}{\sqrt{2}}(|u\bar u \rangle-| d\bar d \rangle),\qquad \left|a^+_0(980)\right\rangle=|u\bar d\rangle,\qquad \left|a^-_0(980)\right\rangle=\left|d\bar u \right\rangle.
\en
While such a $q\bar q$ picture meets with several difficulties. For example, since the $s$ quark is expected to be heavier than the $u(d)$ quark, it is difficult to explain the fact that the strange meson $K^*(800)$ is lighter than the isovector meson $a_0(980)$, and the isosinglet meson $f_0(980)$ has a degenerate mass with $a_0(980)$. Furthermore, the P-wave explanation for these light scalar mesons is also difficult to explain why the $K^*_0(800)$ is lighter than its vector partner $K^*(892)$. In order to resolve these difficulties, Jaffe put forward the tetraquark scenario, where the flavor wave functions of $a_0(980)$ read~\cite{jaffe}
\be
\left|a^0_0(980)\right\rangle=\frac{1}{\sqrt{2}}(|s\bar su \bar{u}\rangle-|s\bar sd \bar{d}\rangle), \qquad
\left|a_0^{+}(980)\right\rangle=|s\bar s u \bar{d}\rangle,\qquad 
\left|a_0^{-}(980)\right\rangle=|s\bar sd\bar{u} \rangle.
\en 
Another important isovector scalar meson $a_0(1450)$ above 1 GeV is also received much attention, which was firstly observed in the $p\bar p$ annihilation into $\pi^0\pi^0\eta$ experiment \cite{amsler}.
Compared to the ambiguous assignment for the light scalar mesons below 1 GeV, $a_0(1450)$ is usually interpreted as a $q\bar q$ ground state.

Heavy-flavored meson decays provide a platform for experiments and different theoretical models to explain the nature of scalar mesons. On the experimental side, BaBar firstly studied $a_0(980)$ in the two-body decays $B\to a_0(980)\pi, a_0(980)K$ \cite{babar} with some upper limits obtained. Furthermore, $a_0(1450)$ was also investigated in the decays $B^0\to a^-_0(1450)(\to\eta\pi^-)\pi^+$ and $B^0\to  a^-_0(1450)(\to\eta\pi^-)K^+$ \cite{babar2}. Later, LHCb measured the branching ratio of the decay $\bar B^0\to J/\Psi a_0(980)(\to K^+K^-)$ as $(4.70\pm3.39)\times10^{-7}$, where the evidence of the $a_0(980)$ was reported with a statistical significance of 3.9 standard deviations~\cite{LHCb1}. As an important supplement for the B-meson decays, the D-meson decays also provide another platform to recognize these two isovector mesons $a_0(980)$ and $a_0(1450)$ through, for example, the semileptonic decays $D^{0(+)}\to a_0^{-(0)}(980)e^+\nu_e$~\cite{bes3} as well as the nonleptonic decays $D^+\to a^0_0(980,1450)(\to K^+K^-)\pi^+$~\cite{cleo}, $D^0\to a^\pm_0(980,1450)(\to K^0_SK^\pm)\pi^\mp$ \cite{LHCb2} and
$D^0\to a^\pm_0(980,1450)(\to \pi^{\pm}\eta)\pi^{\mp}, D^+\to a^{+(0)}_0(980,1450)(\to \pi^{+(0)}\eta)\pi^{0(+)}$~\cite{bes31}. These measurements will provide important information for uncovering the inner structures of $a_0(980)$ and $a_0(1450)$. Recently, some authors have explored the decays $D\to a_0(980)P$ with $P$ denoting a pseudoscalar meson using the topological diagrammatic approach, and their conclusions also support the tetraquark nature of the $a_0(980)$ meson by comparing with the data~\cite{frxu,achasov}.

On the theoretical side, the $B$-meson decays to $a_0(980)$ or $a_0(1450)$ have been widely investigated in the QCD factorization (QCDF)~\cite{chenghy1,chenghy2,chenghy3} and the perturbative QCD (PQCD)~\cite{shenyl,zhangzq1,zhangzq2,zouzt,chaij,zhour} approach. Most of these works assume the two-body framework, that is, $a_0(980)$ and $a_0(1450)$  were considered as simple mesons, while they were regarded as intermediate resonant states in Refs.~\cite{chaij} and \cite{zhour}, where the two-meson DAs $\Phi_{K\bar K(\pi\eta)}$ are introduced to describe these two isovector scalar mesons. The large uncertainties from the DAs of $a_0(980)$ and $a_0(1450)$ used in the QCD-inspired calculations are one of the main factors affecting the accuracy of the theoretical predictions.
We find that the Gegenbauer coefficients used under the three-body framework \cite{chaij} are the same as those calculated under the two-body framework \cite{shenyl,zhangzq1,zhangzq2}. It is noticed that the definitions of these two sets of DAs are different, and hence the corresponding Gegenbauer coefficients should be different from each other. In order to decrease the theoretical uncertainties, we would like to use the Gegenbauer coefficient extracted from the data for the decay $B^0\to J/\Psi a_0(980)(\to K^+K^-)$ \cite{LHCb1} to study the quasi-two-body decays $B\to \pi(K)a_0(980,1450)(\to \pi\eta, K\bar K)$. 
  
The layout of this paper is as follows. The framework of the PQCD approach for the quasi-two-body B meson decays is reviewed in Section II, where besides the kinematic variables for each meson, the S-wave $\pi\eta(K\bar K)$ pair distribution amplitudes up to twist-3 are introduced to describe the contributions from the 
intermediate resonant states $a_0(980)$ and $a_0(1450)$. Then, the analytical formulas for the Feynman diagrams and the total amplitudes for our considered decays are listed.  
The numerical results and discussions are given in
Section III, where the comparisons with the results obtained by other works are also included. The conclusions are presented in the final part.
\section{FRAMEWORK}
 In what follows, we will use the abbreviation $a_0$ and $a_0^{\prime}$ to denote the $a_0(980)$ and $a_0(1450)$ for simplicity. We consider the decay $B \rightarrow \pi K \bar{K}$ as an illustration, where $K \bar{K}$ decayed from $a^{(\prime)}_0$ can be either a neutral or a charged kaon pair.
In the rest frame of the $B$ meson, we define the $B$ meson momentum $p_B$, the scalar meson $a_0$ momentum $p$, the bachelor meson $\pi$ momentum $p_3$, and the momentum of the light quark in each meson $k_i$ as follows
\be
p_B&=&\frac{m_B}{\sqrt{2}}\left(1,1, 0 _T\right),\quad p=\frac{m_B}{\sqrt{2}}\left(1, \eta, 0 _T\right), \quad p_3=\frac{m_B}{\sqrt{2}}\left(0, 1-\eta, 0 _T\right), \\
k_B&=&\frac{m_B}{\sqrt{2}} \left(0, x_B, k _{B T}\right), \quad k=\frac{m_B}{\sqrt{2}}(z, 0, k _T),
\quad k_3= \frac{m_B}{\sqrt{2}}\left(0,(1-\eta) x_3, k _{3 T}\right),
\en
where the ratio $\eta=\omega^2/{m^2_B}$ with the invariant mass squared $\omega^2=p^2$ for the $K\bar K$ pair, $x_B, z$ and $x_3$ are the corresponding momentum fractions. The individual kaon momenta $p_1$ and $p_2$ in the $K \bar{K}$ pair are written as
\be
p_1=\left(\zeta p^{+}, \eta(1-\zeta) p^{+}, \omega \sqrt{\zeta(1-\zeta)}, 0\right), \quad p_2=\left((1-\zeta) p^{+}, \eta \zeta p^{+},-\omega \sqrt{\zeta(1-\zeta)}, 0\right),
\en
with $\zeta$ being the kaon momentum fraction. The three-momenta of the kaon and the bachelor meson $\pi$ in the $K \bar{K}$ center of mass are given by
\be
\left|\vec{p}_1\right|=\frac{\lambda^{1 / 2}\left(\omega^2, m_K^2, m_K^2\right)}{2 \omega}, \quad\left|\vec{p}_3\right|=\frac{\lambda^{1 / 2}\left(m_B^2, m^2_\pi, \omega^2\right)}{2 \omega},
\en
respectively, with the Källén function $\lambda(a, b, c)=a^2+b^2+c^2-2(a b+a c+b c)$. The kinematic expressions for the $\pi \eta$ final state can be obtained from the above with a simple replacement.

In the framework of the PQCD approach for the quasi-two-body decay, the factorization formula of the decay amplitude is given as
\be
\mathcal{A}=\Phi_{B} \otimes H \otimes \Phi^{\text{S-wave}}_{K\bar K(\pi\eta)} \otimes \Phi_{h},
\en
 where $H$ refers to the hard kernel for the $b$ quark decay and can be calculated perturbatively, $h$ represents the bachelor meson ($\pi, K$) and the nonperturbative dynamics are absorbed into the DAs of the S-wave $K\bar K(\pi\eta)$ pair, the initial and the bachelor mesons.  The $B$ meson wave function with an intrinsic $b$ (the conjugate space coordinate to $k_T$ ) dependence is expressed as \cite{H. N. Li}
\be
\Phi_B(x, b)=\frac{i}{\sqrt{2 N_c}}\left[\left(p\!\!\!/_B+m_B\right) \gamma_5 \phi_B(x, b)\right],
\en
where only the contribution of Lorentz structure $\phi_B(x, b)$ is included, since the other Lorentz structure $\bar\phi_B$ is numerically small and has been neglected. For the DA $\phi_B(x,b)$, we adopt the following form \cite{H. N. Li, T. Kurimoto}
\be
\phi_B(x, b)=N x^2(1-x)^2 \exp \left[-\frac{x^2 m_B^2}{2 \omega_b^2}-\frac{\omega_b^2 b^2}{2}\right],
\en
where the shape parameter $\omega_b=0.40 \pm 0.04$ GeV in the numerical calculations, and  the normalization factor $N_B=91.745$ for $\omega_b=0.40$ GeV.

The S-wave $K\bar K(\pi\eta)$ system light-cone distribution amplitudes (LCDAs) are given as \cite{WFWangHNLiWWangandCDLu, CHChenHNLi}
\begin{equation}
	\Phi_{K\bar K(\pi\eta)}(z, \omega^2)=\frac{1}{\sqrt{2 N_c}}\left[p\!\!\!/\phi_0(z, \omega^2)+\omega \phi_s(z, \omega^2)+\omega(v\!\!\!/ n\!\!\!/ -1) \phi_t(z, \omega^2)\right],
\end{equation}
where the dimensionless vectors $n=(1,0, 0 _T)$ and $v=(0,1, 0 _T)$, the twist-2 LCDA $\phi_0(z,\omega^2)$ and the twist-3 LCDAs $\phi_{s,t}(z,s)$ are adopted as follows 
\be
\phi_0(z,\omega^2)&=&\frac{9}{\sqrt{2 N_c}} F_{K \bar{K}(\pi \eta)}(\omega^2) B_1 z(1-z)(1-2 z), \\
\phi_s(z,\omega^2)&=&\frac{1}{2 \sqrt{2 N_c}} F_{K \bar{K}(\pi \eta)}(\omega^2), \\
\phi_t(z,\omega^2)&=&\frac{1}{2 \sqrt{2 N_c}} F_{K \bar{K}(\pi \eta)}(\omega^2)(1-2 z). 
\en
It is noticed that the concerned isovector scalar dimeson systems have similar asymptotic DAs as those for the light scalar meson \cite{ZRuiWFWang,UMeinerWWang}, but we replace the scalar decay constant with the time-like form factor. These formulae are the same as the two-pion LCDAs \cite{WFWangHNLiWWangandCDLu} except for the different Gegenbauer coefficient $B_1$ due to the $SU (3)$ breaking effects. Here $B_1=0.3$ is determined from the data for the decay $B^0 \rightarrow J/\psi a_0\left(\rightarrow K^{+} K^{-}\right)$ \cite{Aaij}. The isovector scalar form factor for the $K\bar{K}(\pi \eta)$ dimeson system from $a_0$ is given as
\be
F_{K \bar{K}(\pi \eta)}(\omega^2)=C_{a_0} \frac{m_{a_0}^2}{m_{a_0}^2-\omega^2-i\left(g_{\pi \eta}^2 \rho_{\pi \eta}+g_{K \bar K}^2 \rho_{K \bar K}\right)},
\en
where the coupling constants $g_{\pi \eta}=0.324$ GeV and $g_{K \bar K}=0.329$ GeV are taken from the Crystal Barrel 
experiment \cite{Abele}. The factors $\rho_{\pi\eta,K\bar K}$ are given by the Lorentz-invariant phase space
\be
\rho_{\pi \eta}&=&\sqrt{\left[1-\left(\frac{m_\eta-m_\pi}{\omega}\right)^2\right]\left[1-\left(\frac{m_\eta+m_\pi}{\omega}\right)^2\right]}, \\
\rho_{K \bar{K}} & =&\frac{1}{2} \sqrt{1-\frac{4 m_{K^{ \pm}}^2}{\omega^2}}+\frac{1}{2} \sqrt{1-\frac{4 m_{K^0}^2}{\omega^2}}.
\en
While the isovector scalar form factor for the $K\bar{K}(\pi \eta)$ dimeson system from $a_0^\prime$ is defined through the Breit-Wigner model,
\be
F_{K \bar{K}(\pi \eta)}(\omega^2)=C_{a^\prime_0}\frac{m_{a^\prime_0}^2}{m_{a^\prime_0}^2-\omega^2-i m_{a^\prime_0} \Gamma(\omega)},
\en
where the energy dependent decay width $\Gamma(\omega)$ is related with the total decay width $\Gamma_{a^\prime_0}$ of $a_0^\prime$ through the formula
\be
\Gamma(\omega)=\Gamma_{a^\prime_0} \frac{\left|\vec{p}_1\right|}{\left|\vec{p}_{10}\right|} \frac{m_{a^\prime_0}}{\omega},
\en
with $m_{a^\prime_0}=1.474$ GeV and $\Gamma_{a^\prime_0}=0.282$ GeV for the $a_0^{\prime}$ resonance \cite{Aaij2}. The $\left|\vec{p}_{10}\right|$ is the value for $\left|\vec{p}_1\right|$ at the resonance peak mass.

For the $K \bar{K}$ pair, the magnitude $|C^{K\bar{K}}_{a^{(\prime)}_0}|$ and phase $\phi_{a^{(\prime)}_0}$ can be obtained by fitting the data
in the isobar model for the decay $D^0\to K^0_SK^-\pi^+$ \cite{Aaij2}, which are given as
\be
\left|C^{K \bar{K}}_{a_0}\right|=1.07, \quad \phi_{a_0}=82^{\circ}, \quad\left|C^{K \bar{K}}_{a_0^{\prime}}\right|=0.43. \quad \phi_{a_0^{\prime}}=-49^{\circ}.
\en
While the experimental measurements of the complex amplitudes for the $\pi \eta$ pair are not yet available, one can roughly estimate their magnitudes $\left|C_{a^{(\prime)}_0}\right|$ through the relation \cite{zhour},
\be
\frac{C_{a_0}^{\pi \eta}}{C_{a_0}^{K \bar{K}}}=\frac{g_{a_0 \pi \eta}}{g_{a_0 K \bar K}},\quad \frac{C_{a^\prime_0}^{\pi \eta}}{C_{a^\prime_0}^{K \bar{K}}}=\frac{g_{a^\prime_0 \pi \eta}}{g_{a^\prime_0 K \bar K}}.
\en
Using the upper ratios of the strong coupling constants for $a_0$ and $a_0^{\prime}$ determined from the Crystal Barrel experiment \cite{Abele}, one can get the magnitude $\left|C_{a^{(\prime)}_0}\right|$ and phase $\phi_{a^{(\prime)}_0}$ for the $\pi\eta$ system, which are given as follows
\be
\left|C^{\pi\eta}_{a_0}\right|=1.05, \quad \phi_{a_0}=82^{\circ}, \quad\left|C^{\pi\eta}_{a_0^{\prime}}\right|=0.46, \quad \phi_{a_0^{\prime}}=-49^{\circ},
\en
where we keep the phase $\phi_{a^{(\prime)}_0}$ the same with those in $K\bar K$ system.

As to the bachelor mesons $K$ and $\pi$, their DAs have been well determined in many papers, and the related parameters up to twist-3 can be found in our previous work \cite{zhaoyc}. 
\section{Perturbative calculation}

\begin{figure}[t]
	\centering
	\includegraphics[width=0.99\textwidth]{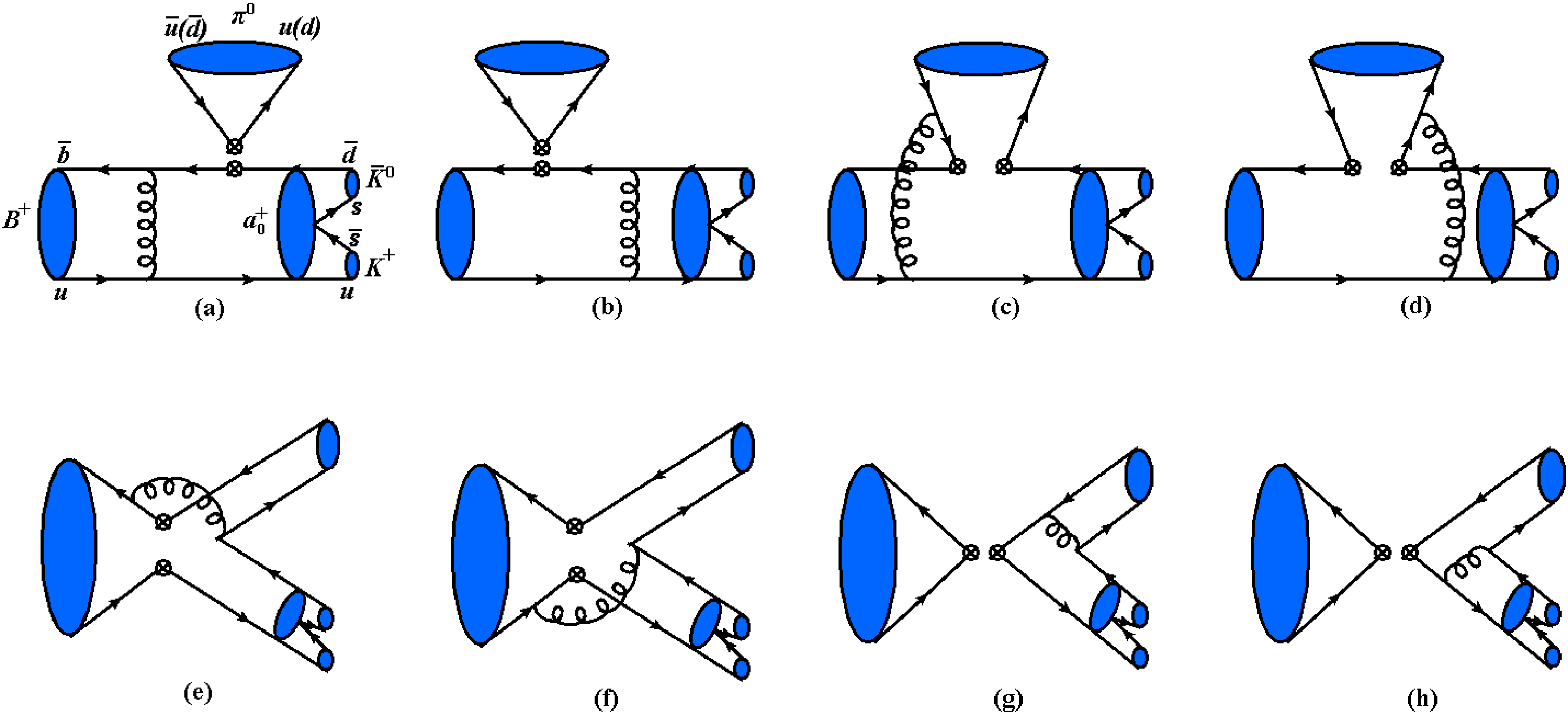}
	\caption{Typical Feynman diagrams for the quasi-two-body decay $B^+\to a_0^{+}(\to \bar K^0 K^+)\pi^0$.}
	\label{a0bark}
\end{figure}

The typical Feynman diagrams at the leading order for the quasi-two-body decay $B^+\to a_0^{+}(\to \bar K^0 K^+)\pi^0$ are shown in Fig. \ref{a0bark}. In the following, we will take this channel as an example to analyze the analytical formulas for each Feynman diagram, which include the contributions from $(V-A)(V-A), (V-A)(V-A)$ and $(S-P)(S+P)$ operators, respectively. The amplitudes from the factorizable emission diagrams shown in Figs.~\ref{a0bark}(a) and \ref{a0bark}(b) can be written as
\be
\mathcal{F}^{LL}_e&=&8\pi
C_Fm^4_{B}\int_0^1 dx_{B} dz\, \int_{0}^{\infty} b_B db_B
b db\, \phi_{B}(x_B,b_B)f_\pi(\eta-1)\left\{\left[(1+z)\phi_0\right.\right.\non && \left.\left.
+(1-2z)\sqrt\eta(\phi_s+\phi_t)\right]
\alpha_s(t_a)S_t(z)h(\alpha_e,\beta_a,b_B,b)\exp[-S_{ab}(t_a)]\right.\non &&\left.\left.
-\sqrt\eta[\sqrt\eta \phi_0-2\phi_s]\alpha_s(t_b)S_t(x_B) h(\alpha_e,\beta_b,b,b_B)\exp[-S_{ab}(t_b)\right]\right\},\\	
\mathcal{F}^{LR}_e &=&-\mathcal{F}^{LL}_e,			\\
\mathcal{F}^{SP}_e&=& 16\pi C_Fm^4_{B}f_\pi r_\pi \int_0^1 dx_{B} dz\, \int_{0}^{\infty} b_B db_Bb db\phi_{B}(x_B,b_B)\left\{\left[\phi^0(\eta(2z-1)-1)\right.\right.\non && \left.\left.
+\sqrt\eta(z\phi^t-(2+z)\phi^s) \right]
\alpha_s(t_a)S_t(z)h(\alpha_e,\beta_a,b_B,b)\exp[-S_{ab}(t_a)]\right.\non&&\left.
+[(2\eta-x_B)\phi^0-2\sqrt\eta(1+\eta-x_B)\phi^s]\right.\non&&\left.\times\alpha_s(t_b)S_t(x_B)h(\alpha_e,\beta_b,b,b_B)\exp[-S_{ab}(t_b)]\right\},
\en
where the mass ratio $r_\pi=m_{0}/M_B$ and $f_\pi$ is the decay constant of $\pi$ meson. The hard function $h_e$, the hard scales $t_{a,b}$, the Sudakov exponent $\exp[-S_{ab}(t)]$ and the threshold
resummation factor $S_t(x)$ are listed in Appendix A. The amplitudes for the nonfactorizable emission diagrams Figs.~\ref{a0bark}(c) and 
\ref{a0bark}(d) are given as
\be
    \mathcal{M}^{LL}_{e}&=&32\pi C_f m_{B}^4/\sqrt{2N_C}\int_0^1 d x_{B} dz dx_{3}
\int_{0}^{\infty} b_B db_B b_3 db_3\,\phi_{B}(x_B,b_B)\phi_\pi^A(x_3)(\eta-1)\non &&
\times\left\{\left[(\eta(1-z-x_3)-1+x_B+x_3)\phi^0+\sqrt{\eta}z(\phi^s-\phi^t)\right]\right.\non && \left.\times \alpha_s(t_{c})h(\beta_c,\alpha_e,b_B,b_3)\exp[-S_{cd}(t_c)]+\left[(z-x_B-(\eta-1)x_3)\phi^0\right.\right.\non && \left.\left. -z\sqrt\eta(\phi^s+\phi^t) \right]\alpha_s(t_d)h(\beta_d,\alpha_e,b_3,b_B)\exp[-S_{cd}(t_d)]\right\}, \label{nfe1}\\
	\mathcal{M}^{LR}_{e}&=&32\pi C_f m_{B}^4r_\pi/\sqrt{2N_C}\int_0^1 d x_{B} dz dx_{3}\int_{0}^{\infty} b_B db_B b_3 db_3\,\phi_{B}(x_B,b_B)\non &&
\times\left\{\left[(\eta(1-x_3)+x_B+x_3-1)(\phi^0+\sqrt\eta(\phi^s-\phi^t))(\phi^P_\pi+\phi^T_\pi)+\sqrt\eta z(\sqrt\eta\phi^0\right.\right.\non &&\left.\left. +\phi^s+\phi^t)(\phi^T_\pi-\phi^P_\pi)\right]\alpha_s(t_{c})h(\beta_c,\alpha_e,b_B,b_3)\exp[-S_{cd}(t_c)]
+\left[((1-\eta)x_3-x_B) \right.\right.\non &&\left.\left.+\sqrt\eta(\phi^s-\phi^t))(\phi^P_\pi-\phi^T_\pi)\right]+\sqrt\eta z(\sqrt\eta\phi^0+\phi^s+\phi^t)(\phi^P_\pi+\phi^T_\pi)\right.\non &&\left.\times \alpha_s(t_d)h(\beta_d,\alpha_e,b_3,b_B)\exp[-S_{cd}(t_d)]\right\},\\
\mathcal{M}^{SP}_{e}&=&-32\pi C_f m_{B}^4/\sqrt{2N_C}\int_0^1 d x_{B} dz dx_{3}
\int_{0}^{\infty} b_B db_B b_3 db_3\,\phi_{B}(x_B,b_B)\phi_\pi^A(x_3)(\eta-1)\non &&
\times\left\{\left[(\eta(1-x_3)+x_B-z+x_3-1)\phi^0+\sqrt\eta z(\phi^s+\phi^t)\right]\right.\non && \left.\times \alpha_s(t_{c})h(\beta_c,\alpha_e,b_B,b_3)\exp[-S_{cd}(t_c)]-[(x_B-z\eta+(\eta-1)x_3)\phi^0\right.\non && \left.+\sqrt\eta z(\phi^s-\phi^t)]
\alpha_s(t_d)h(\beta_d,\alpha_e,b_3,b_B)\exp[-S_{cd}(t_d)]\right\}.
\en
It is noticed that the integration over the intrinsic variable $b$ has been already performed using the Dirac $\delta$ function $\delta(b_B-b)$, leaving only the integration over $b_B$ and $b_3$. The amplitudes from the nonfactorizable annihilation diagrams Figs.~\ref{a0bark}(e) and \ref{a0bark}(f) are written as 
\be
\mathcal{M}^{LL}_{a}&=& 32\pi C_f m_{B}^4/\sqrt{2N_C}\int_0^1 d x_{B} dz dx_{3} \int_{0}^{\infty} b_B db_B b_3 db_3\,\phi_{B}(x_B,b_B)\non && 
\times\left\{\left[(\eta^2(1-z-x_3)+\eta(x_B+2x_3+z-1)-x_B-x_3)\phi^0\phi^A_\pi\right.\right.\non &&\left.\left. +\sqrt\eta r_\pi((x_B+(1-x_3)(\eta-1))(\phi^s-\phi^t)(\phi^P_\pi+\phi^T_\pi)-z(\phi^s+\phi^t)(\phi^P_\pi-\phi^T_\pi)\right.\right.\non &&\left.\left.+4\phi^P_\pi\phi^T_\pi)
\right]\alpha_s(t_{e})h(\beta_e,\alpha_a,b_B,b_3)\exp[-S_{ef}(t_e)]+\left[(\eta^2-1)(z-1)\phi^0\phi^A_\pi\right.\right.\non &&\left.\left.+\sqrt\eta r_\pi((x_B-\eta+(\eta-1)x_3)(\phi^s+\phi^t)(\phi^P_\pi-\phi^T_\pi)+(z-1)(\phi^s-\phi^t)\right.\right.\non && \left.\left.(\phi^P_\pi+\phi^T_\pi))
\right] \alpha_s(t_{f})h(\beta_f,\alpha_a,b_3,b_B)\exp[-S_{ef}(t_f)]\right\},\\
\mathcal{M}^{LR}_{a}&=& 32\pi C_f m_{B}^4/\sqrt{2N_C}\int_0^1 d x_{B} dz dx_{3} \int_{0}^{\infty} b_B db_B b_3 db_3\,\phi_{B}(x_B,b_B)\non && 
\times\left\{\left[\sqrt\eta(z+1)(\eta-1)\phi^0(\phi^s-\phi^t)+r_\pi\phi^0((x_B-2+x_3(1-\eta))(\phi^P_\pi+\phi^T_\pi)
\right.\right.\non && \left.\left. -z\eta(\phi^P_\pi-\phi^T_\pi)+2\eta\phi^T_\pi)\right]\alpha_s(t_{e})h(\beta_e,\alpha_a,b_B,b_3)\exp[-S_{ef}(t_e)]\right.\non &&\left.+\left[
\sqrt\eta(z-1)(1-\eta)\phi^0(\phi^s-\phi^t)+r_\pi\phi^0((x_1-x_3+\eta x_3)(\phi^P_\pi+\phi^T_\pi)\right.\right.\non &&\left.\left.+z\eta (\phi^P_\pi-\phi^T_\pi)-2\eta\phi^P_\pi)\right] \alpha_s(t_{f})h(\beta_f,\alpha_a,b_3,b_B)\exp[-S_{ef}(t_f)]\right\},\\
\mathcal{M}^{SP}_{a}&=& 32\pi C_f m_{B}^4/\sqrt{2N_C}\int_0^1 d x_{B} dz dx_{3} \int_{0}^{\infty} b_B db_B b_3 db_3\,\phi_{B}(x_B,b_B)\non && 
\times\left\{\left[(\eta-1)(z\eta+z-1)\phi^0\phi^A_\pi+\sqrt\eta r_\pi(((1-\eta)(1-x_3)-x_B+1)\right.\right.\non && \left.\left. \times (\phi^s+\phi^t)(\phi^P_\pi-\phi^T_\pi)+z(\phi^s-\phi^t)(\phi^P_\pi+\phi^T_\pi)-4\phi^s\phi^P_\pi)\right]\right.\non && \left. \times\alpha_s(t_{e})h(\beta_e,\alpha_a,b_B,b_3)\exp[-S_{ef}(t_e)]\right.\non && \left.+\left[(1-\eta)(x_B+(\eta-1)x_3+\eta(z-2))\phi^0\phi^A_\pi-r_\pi\sqrt\eta((x_B+(\eta-1)x_3-\eta)\right.\right.\non &&\left.\left.
\times(\phi^s-\phi^t)(\phi^P_\pi+\phi^T_\pi)+(z-1)(\phi^s+\phi^t)(\phi^P_\pi-\phi^T_\pi))
\right]\right.\non && \left.\times \alpha_s(t_{f})h(\beta_f,\alpha_a,b_3,b_B)\exp[-S_{ef}(t_f)]\right\}.
\en
For the factorizable annihilation diagrams Figs.~\ref{a0bark}(g) and \ref{a0bark}(h), the corresponding amplitudes are given as
\be
\mathcal{F}^{LL}_{a}&=& 8\pi C_F m_{B}^4f_B/\sqrt{2N_C}\int_0^1  dz dx_{3} \int_{0}^{\infty} b db b_3 db_3 \times\left\{[(1-\eta)(z-1)\phi^0\phi^A_\pi \right.\non && \left.
-2r_\pi \sqrt\eta\phi^P_\pi((z-2)\phi^s-z\phi^t)] \alpha_s(t_g)S_t(x_3)h(\alpha_a,\beta_g,b,b_3)\exp[-S_{gh}(t_g)]\right.\non &&\left.+\left[(1-\eta)((1-\eta)x_3+\eta)\phi^0\phi^A_\pi+2\sqrt\eta r_\pi \phi^s(((\eta-1)x_3-\eta)(\phi^P_\pi+\phi^T_\pi)\right.\right.\non &&\left.\left.
-\phi^P_\pi+\phi^T_\pi)\right]\alpha_s(t_h)S_t(z)h(\alpha_a,\beta_h,b_3,b)\exp[-S_{gh}(t_f)]\right\},\\
\mathcal{F}^{LR}_{a}&=&-\mathcal{F}^{LL}_{a}, \\
\mathcal{F}^{SP}_{a}&=& 16\pi C_Fm_{B}^4f_B /\sqrt{2N_C}\int_0^1 d z  dx_{3} \int_{0}^{\infty} b db b_3 db_3\non && 
\times\left\{\left[2r_\pi(\eta(z-1)-1)\phi^0\phi^P_\pi+\sqrt\eta(z-1)(\eta-1)\phi^A_\pi(\phi^s+\phi^t)\right]\right.\non && \left.\times \alpha_s(t_g)S_t(x_3)h(\alpha_a,\beta_g,b,b_3)\exp[-S_{gh}(t_g)]+\left[2\sqrt\eta(1-\eta)\phi^s\phi^A_\pi-r_\pi x_3(1-\eta)\right.\right.\non && \left.\left.\times\phi^0(\phi^P_\pi-\phi^T_\pi)-2\eta r_\pi \phi^0\phi^P_\pi\right]\alpha_s(t_h)S_t(z)h(\alpha_a,\beta_h,b_3,b)\exp[-S_{gh}(t_f)]\right\}.
\en
Certainly, if one replaces the positions of the final states in Fig.~\ref{a0bark}, the newly obtained Feynman diagrams might also contribute to some channels and the analytical formulas
for the corresponding amplitudes are collected in Appendix B. The total amplitudes for the considered decays combined from the different Feynman diagrams can be found in Appendix C.  
\section{Numerical results and discussions}

The input parameters used in the numerical calculations are listed in Table \ref{parame}. 
\begin{table}[h!]
\caption{The values of the masses and the decay constants of the related mesons (in units of GeV) and the Wolfenstein parameters \cite{pdg}.}
\begin{center}
\begin{tabular}{@{\extracolsep{1.2cm}}cccc}
\hline\hline
$m_{B^0} = 5.280$ & $m_{B^{\pm}} = 5.279$& $m_{\pi^{\pm}} = 0.140$ & $m_{\pi^0} = 0.135$\\ 
\hline
$m_{K^{\pm}} = 0.494$ & $m_{K^0} = 0.498$ & $m_{\eta} = 0.548$ &$f_B = 0.210$\\
\hline
$f_{\pi} = 0.130$ &$f_{K} = 0.16$  & $m_{a_0} = 0.980$ & $m_{a_0'} = 1.474$\\
\hline
$\lambda = 0.22650 \pm 0.00048$ & $A = 0.790^{+0.017}_{-0.012}$ &$\bar{\rho} = 0.141^{+0.016}_{-0.017}$& $\bar{\eta} = 0.357 \pm 0.011$ \\
\hline\hline
\end{tabular}
\end{center}
\label{parame}
\end{table}

We present the branching ratios of the quasi-two-body decays $B \rightarrow a_0(\rightarrow \pi \eta, K \bar{K}) h$ and $B \rightarrow a_0^{\prime}(\rightarrow \pi \eta, K \bar{K}) h$ decays in Tables \ref{tab2} and \ref{tab3}, respectively. Besides the results for these quasi-two-body decays listed in the second column, we also give the branching ratios of their corresponding two-body decays by using the narrow width approximation, which are listed in the 4th column. For comparison, the corresponding results given in Ref. \cite{chaij} are listed in the 3rd and 5th columns, respectively. The predictions calculated under two-body framework using the QCDF \cite{zhenghaiyang} and the PQCD \cite{shenyl, zhangzhiqing} approaches are listed in the  6th and 7th columns, respectively.  The available data \cite{pdg} are presented in the last column. 
 For each entry, the first uncertainty is from the shape parameter in the B meson wave function $\omega=0.4\pm0.4$ GeV, and the second error comes from the Gegenbauer coefficient $B_1=0.3\pm0.1$ in the twist-2 DAs of $K\bar K (\pi\eta)$ pair.
\begin{table}[h!]
	\caption{The PQCD predictions for the branching ratios (in units of $10^{-6}$) of the decays $B \to a_0(\to \pi\eta, K\bar K) h$. For comparison, we also list the previous theoretical results and the present data.}
	\begin{center}
		\scalebox{0.86}{
			\begin{tabular}{|c|c|c|c|c|c|c|c|}
				\hline
				\multirow{2}{*}{Decay modes}&\multicolumn {2}{|c|}{Quasi-two-body}&\multicolumn {4}{|c|}{Two-body}&\multirow{2}{*}{Data \cite{pdg}} \\
				\cline{2-7}
				~&this work&Ref. \cite{chaij}&this work&Ref. \cite{chaij}&Ref. \cite{zhenghaiyang}&Refs. \cite{shenyl,zhangzhiqing}& \\
				\hline
				$B^{+}\to a_0^+(\rightarrow\pi^+\eta)K^0$  &$0.44^{+0.14 +0.19}_{-0.07-0.00}$&$0.94^{+0.85}_{-0.51}$&$0.55^{+0.18+0.24}_{-0.09-0.00}$&$1.35^{+1.21}_{-0.72}$&\multirow{2}{*}{$0.08^{+2.20}_{-0.11}$}& \multirow{2}{*}{$6.9^{+2.4}_{-2.1}$}&\multirow{2}{*}{$<3.9$}\\
				$B^{+}\to a_0^+(\rightarrow K^+\bar K^0)K^0$ &$0.15^{+0.04+0.11}_{-0.03-0.03}$&$0.26^{+0.16}_{-0.10}$& & & &&\\
				\hline
				$B^{+}\to a_0^0(\rightarrow\pi^0\eta)K^+$ &$0.73^{+0.06+0.52}_{-0.03-0.19}$&$1.06^{+0.59}_{-0.42}$&$0.91^{+0.08+0.65}_{-0.04-0.24}$&$1.51^{+0.85}_{-0.61}$&\multirow{2}{*}{$0.34^{+1.12}_{-0.16}$}&\multirow{2}{*}{$3.5^{+1.2}_{-1.2}$} &\multirow{2}{*}{$<2.5$}\\
				$B^{+}\to a_0^0(\rightarrow K^+K^-)K^+$ &$0.12^{+0.01+0.10}_{-0.01-0.04}$&$0.11^{+0.06}_{-0.04}$& & & & &\\
				\hline
				$B^0\to a_0^0(\rightarrow\pi^0\eta)K^0$  &$0.27^{+0.12+0.07}_{-0.06-0.00}$&$1.36^{+0.48}_{-0.56}$&$0.34^{+0.15+0.09}_{-0.08-0.00}$&$1.95^{+0.68}_{-0.79}$&\multirow{2}{*}{$0.05^{+0.91}_{-0.05}$} & \multirow{2}{*}{$4.7^{+1.4}_{-1.5}$} &\multirow{2}{*}{$<7.8$}\\
				$B^0\to a_0^0(\rightarrow K^+K^-)K^0$ &$0.05^{+0.02+0.03}_{-0.01-0.01}$&$0.11^{+0.05}_{-0.02}$& & & & &\\
				\hline
				$B^0\to a_0^-(\rightarrow\pi^-\eta)K^+$ &$1.77^{+0.36+1.23}_{-0.24-0.48}$&$4.51^{+1.82}_{-1.71}$&$2.21^{+0.45+1.54}_{-0.30-0.60}$&$6.44^{+2.67}_{-2.49}$&\multirow{2}{*}{$0.34^{+2.35}_{-0.14}$}&\multirow{2}{*}{$9.7^{+3.3}_{-2.8}$} &\multirow{2}{*}{$<1.9$}\\	
				$B^0\to a_0^-(\rightarrow K^- K^0)K^+$ &$0.57^{+0.08+0.40}_{-0.07-0.21}$&$0.99^{+0.40}_{-0.34}$& & & & &\\	
				\hline
				$B^{+}\to a_0^+(\rightarrow\pi^+\eta)\pi^0$  &$0.44^{+0.00 +0.10}_{-0.12 -0.18}$&$0.37^{+0.15}_{-0.09}$&$0.55^{+0.00+0.13}_{-0.15-0.23}$&$0.52^{+0.21}_{-0.12}$&\multirow{2}{*}{$0.70^{+0.32}_{-0.23}$}&\multirow{2}{*}{$0.41^{+0.00}_{-0.23}$}&\multirow{2}{*}{$<1.4$}\\
				$B^{+}\to a_0^+(\rightarrow K^+\bar K^0)\pi^0$  &$0.08^{+0.00 +0.02}_{-0.03 -0.04}$&$0.08^{+0.03}_{-0.03}$& & & & &\\
				\hline
				$B^{+}\to a_0^0(\rightarrow\pi^0\eta)\pi^+$  &$9.90^{+0.40 +5.50}_{-0.32 -3.33}$&$2.41^{+0.98}_{-0.69}$&$12.4^{+0.5+0.7}_{-0.4-4.2}$&$3.44^{+1.40}_{-0.98}$&\multirow{2}{*}{$4.9^{+1.4}_{-1.3}$}&\multirow{2}{*}{$2.8^{+0.0}_{-1.3}$}&\multirow{2}{*}{$<5.8$}\\
				$B^{+}\to a_0^0(\rightarrow K^+K^-)\pi^+$  &$1.06^{+0.02 +0.56}_{-0.00 -0.34}$&$0.33^{+0.13}_{-0.09}$& & & & &\\
				\hline
				$B^0\to a_0^0(\rightarrow\pi^0\eta)\pi^0$  &$0.33^{+0.03 +0.06}_{-0.14 -0.13}$&$0.33^{+0.10}_{-0.08}$&$0.41^{+0.04+0.08}_{-0.18-0.16}$&$0.47^{+0.14}_{-0.11}$&\multirow{2}{*}{$1.0^{+0.5}_{-0.3}$}&\multirow{2}{*}{$0.51^{+0.12}_{-0.11}$}&\multirow{2}{*}{$-$}\\
				$B^0\to a_0^0(\rightarrow K^+\bar K^0)\pi^0$  &$0.04^{+0.00 +0.00}_{-0.02 -0.02}$&$0.04^{+0.02}_{-0.01}$& & & & &\\
				\hline
				$B^0\to a_0^-(\rightarrow\pi^-\eta)\pi^+$ &$21.7^{+1.60 +11.7}_{-1.27 -7.00}$ &$14.8^{+5.9}_{-4.1}$&$27.1^{+2.0+14.6}_{-1.6-8.8}$&$21.1^{+8.2}_{-6.0}$&\multirow{2}{*}{$5.3^{+1.7}_{-1.4}$}&\multirow{2}{*}{$0.86^{+0.17}_{-0.17}$}&\multirow{2}{*}{$-$}\\	
				$ B^0\to a_0^-(\rightarrow K^-K^0)\pi^+$ &$4.43^{+0.15 +2.17}_{-0.14 -1.38}$&$3.48^{+1.37}_{-0.96}$& & & & &\\	
				\hline
				$ B^0\to a_0^+(\rightarrow\pi^+\eta)\pi^-$ &$0.22^{+0.10 +0.02}_{-0.07 -0.02}$&$0.67^{+0.25}_{-0.17}$&$0.27^{+0.13+0.03}_{-0.09-0.03}$&$0.95^{+0.35}_{-0.24}$&\multirow{2}{*}{$0.58^{+0.65}_{-0.25}$}&\multirow{2}{*}{$0.51^{+0.12}_{-0.12}$}&\multirow{2}{*}{$-$}\\	
				$B^0\to a_0^+(\rightarrow K^+\bar K^0)\pi^-$ &$0.04^{+0.02 +0.00}_{-0.01 -0.00}$&$0.17^{+0.06}_{-0.04}$& & & & &\\	
				\hline
			\end{tabular}\label{tab2}}
	\end{center}
\end{table}

Under the narrow width approximation, one can extract the branching ratio of the two-body decay $B \to a^{(\prime)}_0 h$ from that of the quasi-two-body one $B \to a^{(\prime)}_0(\to \pi \eta, K \bar{K}) h$, that is
\be
\mathcal{B}r(B \to a^{(\prime)}_0(\to \pi \eta, K \bar{K}) h)=\mathcal{B}r(B \to a^{(\prime)}_0 h)\cdot \mathcal{B}r(a^{(\prime)}_0\to \pi \eta, K\bar K),
\en 
where the branching ratios of the strong decays are given as $\mathcal{B}r(a_0 \to \pi\eta)=0.8, \mathcal{B}r(a_0^\prime \to \pi\eta)=0.093$ and $\mathcal{B}r(a_0^\prime \to K\bar K)=0.082$ \cite{pdg}. For the decay $a_0\to K \bar K$ is severely constrained by the phase space,  the corresponding branching ratio is only about $0.12$ obtained from $\Gamma(a_0 \to K\bar K)=0.009$ MeV and $\Gamma_{a_0}=0.075$ MeV. So the branching fractions of the quasi-two-body channels through the strong decay $a_0 \rightarrow \pi \eta$ are much larger than  those through the strong decay $a_0 \rightarrow K \bar{K}$.
It is noticed that
the narrow width approximation cannot be used to the decay modes with the strong decay $a_0\to K\bar K$ involved, because the mass of $a_0(980)$ is below the $K\bar K$ threshold. Therefore, one can only derive the branching ratios of the two-body decays $B\to a_0h$ from the corresponding quasi-two-body channels involving the decay $a_0\to \pi\eta$, which are shown in Table \ref{tab2}. Compared with the previous PQCD calculations \cite{chaij,shenyl}, our predictions for the branching ratios of the decays $B\to a_0 K$ are closer to the QCDF results \cite{zhenghaiyang}, which indicate that the Gegenbauer coefficient $B_1=0.3$ determined from the data might be more reasonable. As to the decays $B\to a_0\pi$, it is strange that the branching ratios of the decays $B^+\to a_0^+\pi^0$ and $B^0\to a^0_0\pi^0, a^+_0\pi^-$ can be comparable with the QCDF \cite{zhenghaiyang} and the previous PQCD results no matter under the three-body \cite{chaij} or the two-body framework \cite{zhangzhiqing} , while those for the decays
$B^+\to a^0_0\pi^+$ and $B^0\to  a^-_0\pi^+$ become much larger than the QCDF \cite{zhenghaiyang} and the PQCD \cite{zhangzhiqing} results obtained under the two-body framework. The reason of causing the large branching ratios of the decays $B^+\to a^0_0(\to \pi^0\eta)\pi^+$  and $ B^0\to a^-_0(\to \pi^-\eta)\pi^+$ should be same, while it seems to occur only in the decay $B^0\to a^-_0(\to \pi^-\eta)\pi^+$ in Ref. \cite{chaij}. The completely similar situation also appears in those decays with $a_0$ replaced by $a^\prime_0$, where we will discuss it in detail later. Compared with the present data, one can find that our predictions for most the decays $B\to a_0K$ are below the experimental upper limits except for that of the channel $ B^0\to a^-_0K^+$, which is slightly larger than the experimental upper limit. Certainly, the previous PQCD results for this channel given in Refs. \cite{chaij, shenyl} are much larger than this upper limit. More seriously, all the PQCD calculations under the two-body framework \cite{shenyl} are larger than the experimental upper limits. In view of this point, the three-body framework should
be more appropriate for the decays $B\to a_0 K$, where $a_0$ is described with the $K \bar K(\pi\eta)$ two-meson DAs normalized to the time-like form factor.  It is necessary to accurately measure the branching ratios of these $B\to a_0 K$ decays especially for the channel $ B^0\to a^-_0K^+$ on future experiments, which are meaningful and helpful to probe the inner structure of $a_0$. For another two important decays $B^+\to a^0_0\pi^+$ and $B^0\to a^-_0\pi^+$, the branching ratio of the former is larger than the present experimental upper limit, while the branching fractions of the latter obtained from the three-body framework display significant tensions with those from the two-body one. In order to clarify these divergences, it is urgent to perform an accurate measurement for these two decays in experiments. 

\begin{table}
	\caption{The same as table I, but for the $B \to a_0^{\prime}(\to K \bar{K}, \pi\eta) h$ decays.}
	\begin{center}
		\begin{tabular}{|c|c|c|c|c|c|c|}
			\hline
			\multirow{2}{*}{Decay modes}&\multicolumn {2}{|c|}{Quasi-two-body}&\multicolumn {3}{|c|}{Two-body} &\multirow{2}{*}{Data \cite{pdg}} \\
			\cline{2-6}
			~&this work&Ref\cite{chaij}&this work&Ref.\cite{chaij}&Ref.\cite{zhenghaiyang}& \\
			\hline
			$B^{+}\to a_0^{\prime+}(\rightarrow\pi^+\eta) K^0$  &$0.51^{+0.10 +0.41}_{-0.07 -0.20}$&$1.51^{+0.79}_{-0.62}$&$5.53^{+1.08+4.41}_{-0.75-2.15}$&$16.3^{+8.4}_{-6.5}$ &\multirow{2}{*}{$4.2^{+18.8}_{-4.8}$}&  \\
			$B^{+}\to a_0^{\prime+}(\rightarrow K^+\bar K^0) K^0$  &$0.48^{+0.08 +0.35}_{-0.08 -0.20}$&$1.29^{+0.68}_{-0.52}$&$5.82^{+0.98+4.27}_{-0.98-2.44}$&$15.8^{+4.6}_{-3.9}$& &     \\
			\hline
			$B^{+}\to a_0^{\prime0}(\rightarrow\pi^0\eta)K^+$  &$0.66^{+0.04 +0.44}_{-0.03-0.26}$&$1.13^{+0.56}_{-0.43}$&$7.10^{+0.43+4.73}_{-0.32-2.80}$&$12.2^{+5.9}_{-4.7}$&\multirow{2}{*}{$2.2^{+8.1}_{-2.2}$} &   \\
			$B^{+}\to a_0^{\prime0}(\rightarrow K^+K^-)K^+$  &$0.29^{+0.02 +0.20}_{-0.01 -0.11}$&$0.50^{+0.23}_{-0.20}$&$7.18^{+0.48+4.88}_{-0.24-2.68}$&$12.3^{+3.9}_{-3.1}$& &    \\
			\hline
			$B^0\to a_0^{\prime0}(\rightarrow\pi^0\eta)K^0$  &$0.32^{+0.07 +0.21}_{-0.06 -0.12}$&$0.78^{+0.40}_{-0.29}$&$3.41^{+0.75+2.26}_{-0.65-1.29}$&$8.34^{+4.29}_{-3.96}$ &\multirow{2}{*}{$1.9^{+7.8}_{-2.2}$}&   \\
			$B^0\to a_0^{\prime0}(\rightarrow K^+K^-)K^0$  &$0.14^{+0.03 +0.10}_{-0.02 -0.05}$&$0.33^{+0.17}_{-0.13}$&$3.52^{+0.74+2.44}_{-0.48-1.22}$&$8.10^{+4.05}_{-3.01}$ & &  \\
			\hline
			$B^0\to a_0^{\prime-}(\rightarrow\pi^-\eta)K^+$ &$1.52^{+0.18 +0.96}_{-0.12 -0.54}$&$3.39^{+1.14}_{-1.09}$&$16.3^{+1.9+10.3}_{-1.3-5.8}$&$36.5^{+14.2}_{-10.8}$&\multirow{2}{*}{$3.5^{+17.5}_{-3.9}$} & \multirow{2}{*}{$<4.7$}  \\
			$B^0\to a_0^{\prime-}(\rightarrow K^-K^0)K^+$ &$1.40^{+0.13 +0.83}_{-0.13 -0.51}$&$2.93^{+1.16}_{-0.96}$&$17.1^{+1.6+10.1}_{-1.6-6.2}$&$35.7^{+11.6}_{-10.6}$& &     \\
			\hline
			$B^{+}\to a_0^{\prime+}(\rightarrow\pi^+\eta)\pi^0$  &$0.11^{+0.00 +0.01}_{-0.06 -0.07}$&$0.12^{+0.05}_{-0.03}$&$1.14^{+0.00+0.11}_{-0.65-0.75}$&$1.24^{+0.54}_{-0.33}$&\multirow{2}{*}{$2.1^{+1.1}_{-0.8}$} &    \\
			$B^{+}\to a_0^{\prime+}(\rightarrow K^+\bar K^0)\pi^0$  &$0.10^{+0.00 +0.01}_{-0.05 -0.06}$&$0.10^{+0.04}_{-0.03}$&$1.19^{+0.00+0.12}_{-0.61-0.73}$&$1.24^{+0.52}_{-0.35}$& &    \\
			\hline
			$B^{+}\to a_0^{\prime0}(\rightarrow\pi^0\eta)\pi^+$  &$3.46^{+0.17 +1.50}_{-0.17 -1.03}$ &$0.56^{+0.28}_{-0.18}$&$37.2^{+1.8+16.1}_{-1.8-11.1}$&$6.01^{+3.10}_{-1.98}$  &\multirow{2}{*}{$5.1^{+1.8}_{-1.7}$} &    \\
			$B^{+}\to a_0^{\prime0}(\rightarrow K^+K^-)\pi^+$  &$1.52^{+0.09 +0.67}_{-0.07 -0.44}$ &$0.25^{+0.12}_{-0.08}$&$37.1^{+2.2+16.3}_{-1.7-10.7}$&$6.07^{+2.99}_{-2.02}$& &     \\
			\hline
			$B^0\to a_0^{\prime0}(\rightarrow\pi^0\eta)\pi^0$  &$0.15^{+0.00 +0.01}_{-0.06 -0.07}$&$0.10^{+0.04}_{-0.03}$&$1.56^{+0.00+0.11}_{-0.65-0.75}$&$1.10^{+0.37}_{-0.36}$ &\multirow{2}{*}{$3.3^{+3.1}_{-1.7}$} &   \\
			$B^0\to a_0^{\prime0}(\rightarrow K^+K^-)\pi^0$  &$0.06^{+0.00 +0.01}_{-0.02 -0.02}$&$0.05^{+0.01}_{-0.01}$&$1.58^{+0.00+0.24}_{-0.48-0.48}$&$1.07^{+0.38}_{-0.34}$ & &    \\
			\hline
			$B^0\to a_0^{\prime-}(\rightarrow\pi^-\eta)\pi^+$ &$7.70^{+0.19 +3.30}_{-0.00 -2.05}$ &$4.15^{+1.59}_{-1.13}$&$82.9^{+2.0+35.5}_{-0.0-22.0}$&$44.6^{+17.1}_{-12.3}$ &\multirow{2}{*}{$2.5^{+3.8}_{-1.0}$} &     \\
			$B^0\to a_0^{\prime-}(\rightarrow K^-K^0)\pi^+$ &$6.92^{+0.02 +2.78}_{-0.00 -1.92}$ &$3.61^{+1.37}_{-0.99}$&$84.4^{+0.2+33.9}_{-0.0-23.4}$&$44.0^{+16.9}_{-12.0}$& &      \\
			\hline
			$B^0\to a_0^{\prime+}(\rightarrow\pi^+\eta)\pi^-$ &$0.054^{+0.025 +0.058}_{-0.017 -0.004}$&$0.14^{+0.05}_{-0.03}$&$0.58^{+0.32+0.11}_{-0.11-0.00}$&$1.51^{+0.59}_{-0.39}$ &\multirow{2}{*}{$0.74^{+2.90}_{-0.60}$} &    \\
			$B^0\to a_0^{\prime+}(\rightarrow K^+\bar K^0)\pi^-$ &$0.049^{+0.021 +0.004}_{-0.016 -0.004}$&$0.13^{+0.05}_{-0.03}$&$0.59^{+0.24+0.00}_{-0.24-0.12}$&$1.56^{+0.60}_{-0.39}$& &     \\
			\hline
		\end{tabular}\label{tab3}
	\end{center}
\end{table}

The branching ratios of the strong decays $a_0^\prime \to \pi\eta$ and $a_0^\prime \to K\bar K$ are close to each other, so the quasi-two-body decays with
charged scalar meson $a^{\prime\pm}_0$ involved, namely $B\to a^{\prime\pm}_0(\to \pi^{\prime\pm}\eta)h$ and $B\to a^{\prime\pm}_0(\to K\bar K)h$, have roughly equal branching ratios. While according to the isospin relation $\mathcal{B}r(a_0^{\prime0} \to K^+K^-)=\mathcal{B}r(a_0^{\prime0} \to K^0\bar K^0)=\mathcal{B}r(a^{\prime0}_0\to K\bar K)/2\approx \mathcal{B}r(a_0^{\prime0} \to \pi^0\eta)/2$, the branching ratios of the quasi-two-body decays involving the decay $a^{\prime0}_0\to \pi^0\eta$ are about
two times those of the quasi-two-body decays associated with $a^{\prime0}_0\to K^+K^-$ decay. Compared with the two-body results given in Ref. \cite{chaij}, our predictions for the decays $B\to a_0^\prime K$ 
are also closer to the QCDF calculations, which once again indicate that the Gegenbauer coefficient $B_1=0.3$ determined from the data is more suitable. It is strange that for  the decays $B \to a_0^{\prime}(\to K \bar{K}, \pi\eta) h$ with $h=\pi^0, \pi^-$, their branching ratios given by these two sets of Gegenbauer coefficient values\footnote{One group is $B_1=0.3$ used in our work, the other group is $B_1=-0.58, B_3=-0.49$ by treating $a^\prime_0$ as a lowest-lying $q\bar q$ state (so called scenario II) used in Refs. \cite{chaij, zhenghaiyang}.} are comparable or close to each other, while for the decays $B \to a_0^{\prime}(\to K \bar{K}, \pi\eta) h$ with $h=\pi^+$, the differences of the branching ratios are quite notable, which are shown in Table \ref{tab3}. We have the following arguments: For the decays $ B^0\to a^{\prime-}(\to K \bar{K}, \pi\eta)\pi^+$ and $B^+\to a^{\prime0}(\to K \bar{K}, \pi\eta)\pi^+$, the dominant contributions come from the factorizable emission amplitude $\mathcal{F}^{LL}_e$, where
the LCDAs of the $\pi\eta (K\bar K)$ pair induce unexpected large branching ratios when confronting with the large Wilson coefficient $a_1=C_2+C_1/3$. If we turn off the amplitude $\mathcal{F}^{LL}_e$, the two-body decay results will become 
$Br( B^0\to a^{\prime-}\pi^+)=0.81 \times 10^{-6}$ and $Br(B^+\to a^{\prime 0}\pi^+)=1.7 \times 10^{-6}$. Both of them are smaller than the calculations given by the QCDF approach, $Br( B^0\to a^{\prime-}\pi^+)=(2.5^{+3.8}_{-1.0})\times10^{-6}$ and $Br(B^+\to a^{\prime 0}\pi^+)=(5.1^{+1.8}_{-1.7}) \times 10^{-6}$, respectively. It is similar to these decays with the hadron $a_0^\prime$ replaced by $a_0$. The branching ratios of the decays $ B^0\to a^-_0\pi^+$
and $B^+\to a^0_0\pi^+$ will reduce to $0.94\times10^{-6}$ and $0.69\times10^{-6}$, respectively, with turning off the amplitude $\mathcal{F}^{LL}_e$. If the branching ratios for these two decays $B^0\to a^{-}\pi^+$ and $B^+\to a^{0}\pi^+$ measured by the future experiments are at the order of $10^{-7}$ or even smaller, it is possible that the values of amplitude $\mathcal{F}^{LL}_e$ are overestimated and thereby the simple quark-antiquark picture for the hadron $a_0$ is not appropriate. Certainly, the present Gegenbauer coefficient abstracted from data is very rough and the higher order 
Gegenbauer coefficients are still unavailable. We expect that more accurate Gegenbauer coefficient values for the $\pi\eta (K\bar K)$ pair DAs can be available in the future.

\begin{figure}[ht]
	\centering
	\subfigure[]{\includegraphics[width=0.30\textwidth]{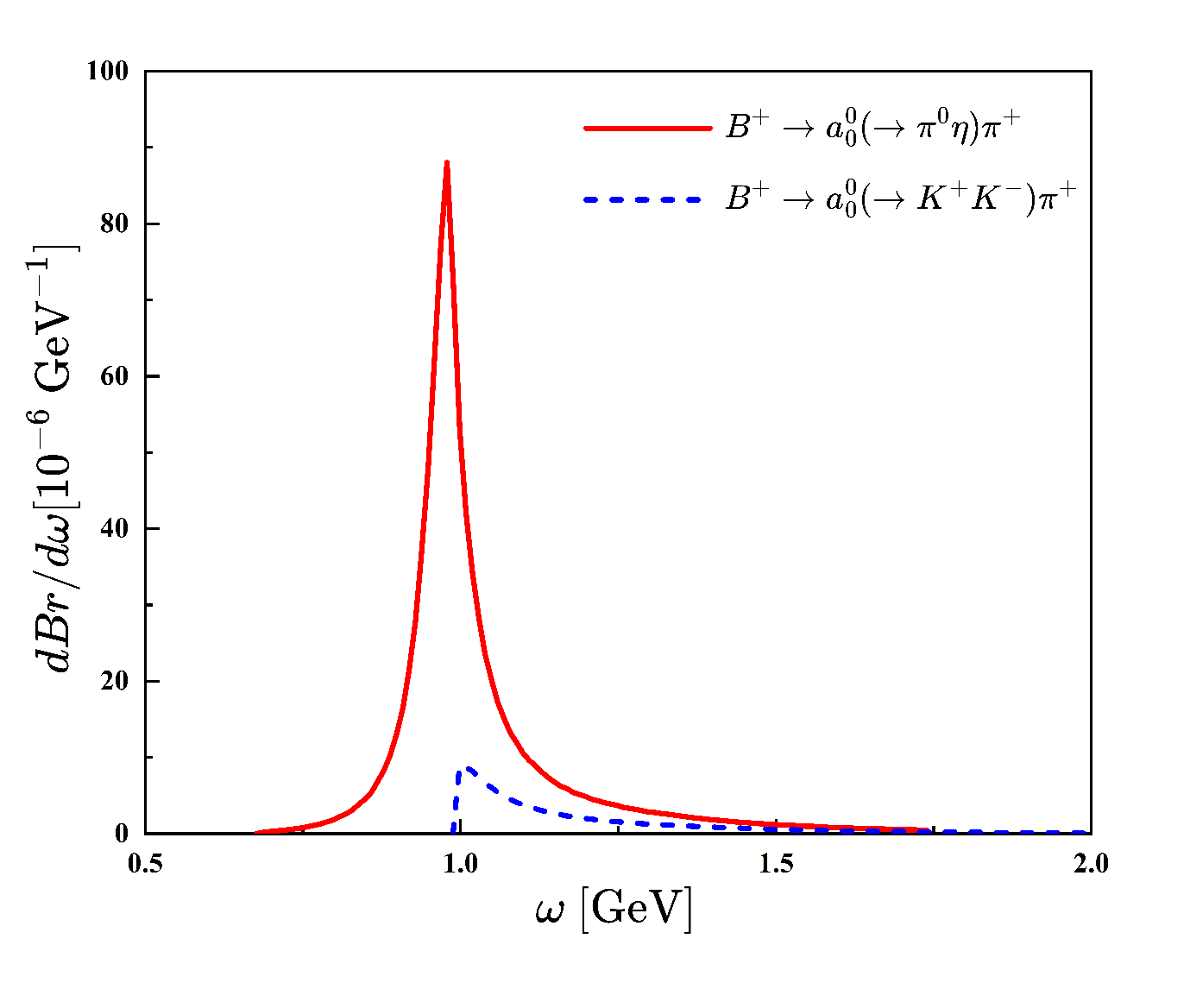}\quad}
	\subfigure[]{\includegraphics[width=0.3\textwidth]{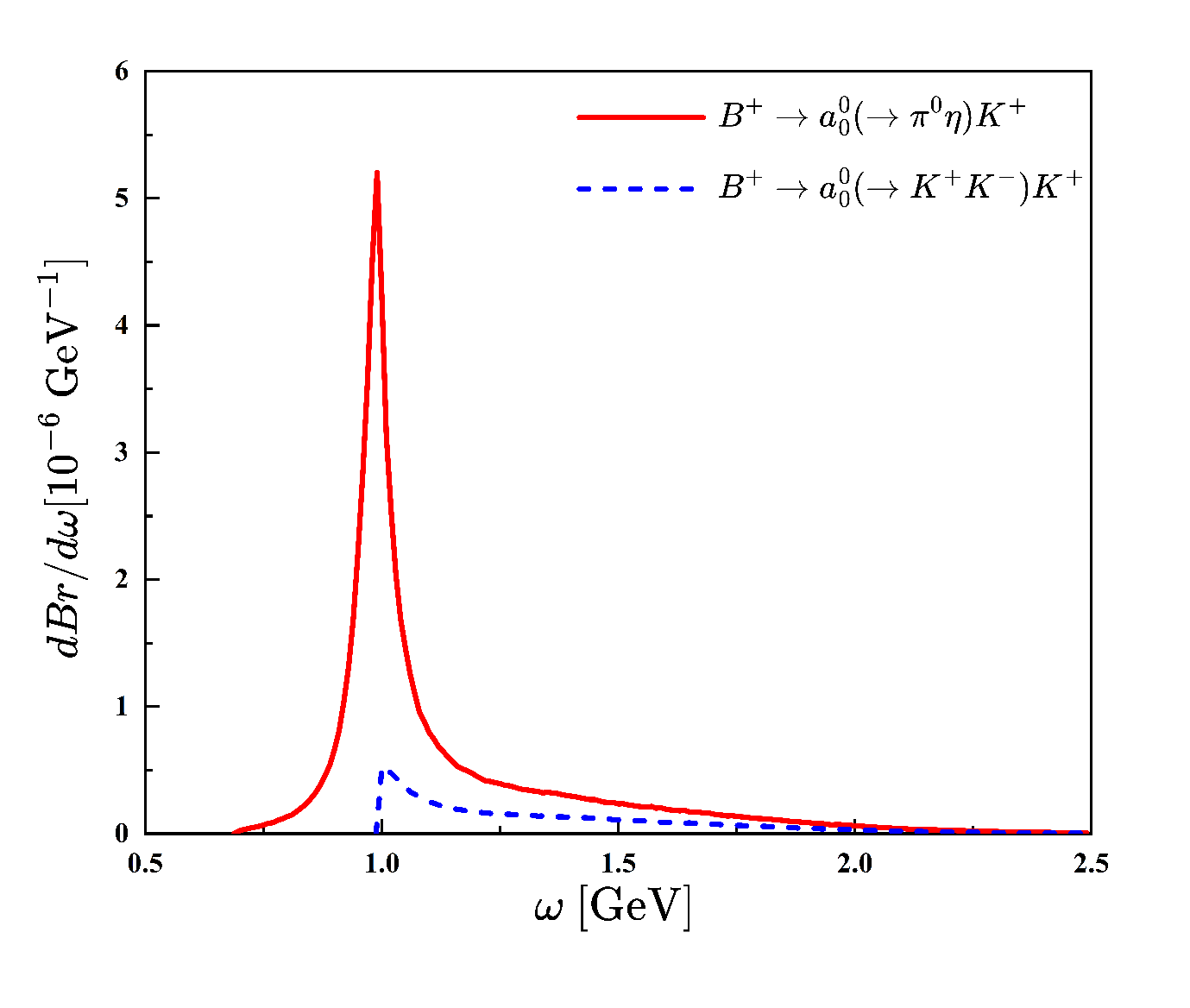}\quad}
	\subfigure[]{\includegraphics[width=0.3\textwidth]{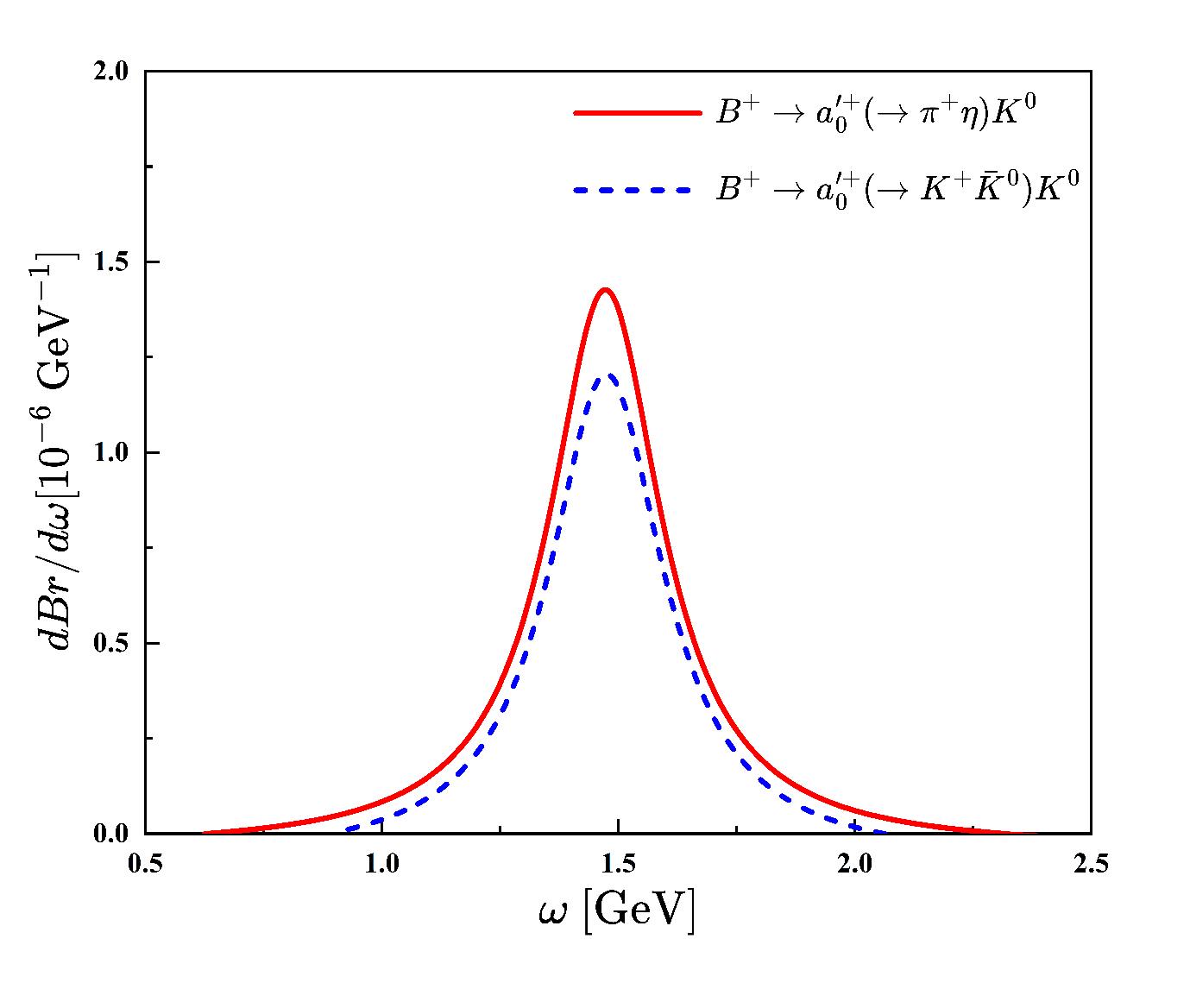}\quad}\\
	\subfigure[]{\includegraphics[width=0.3\textwidth]{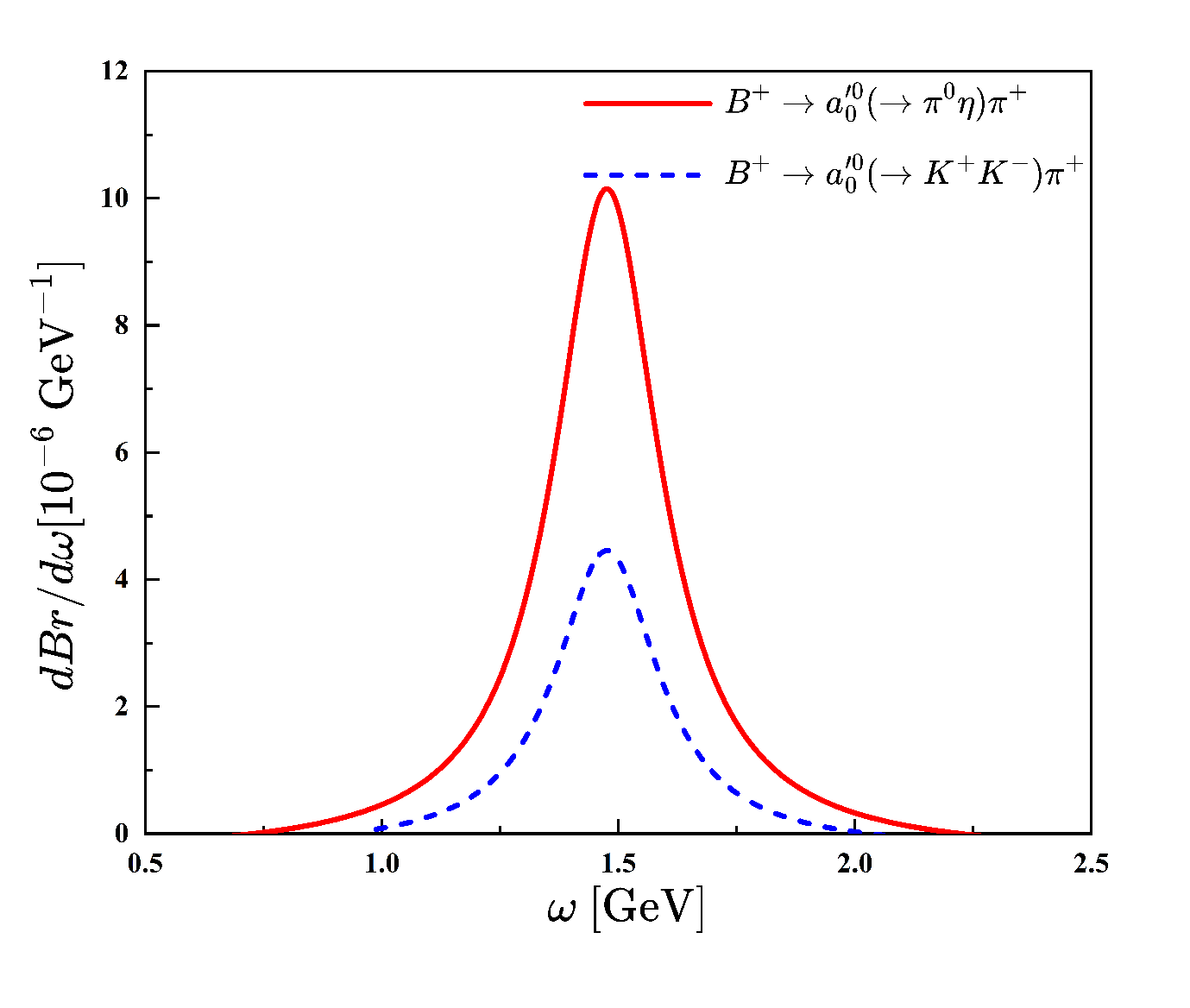}}
	\subfigure[]{\includegraphics[width=0.3\textwidth]{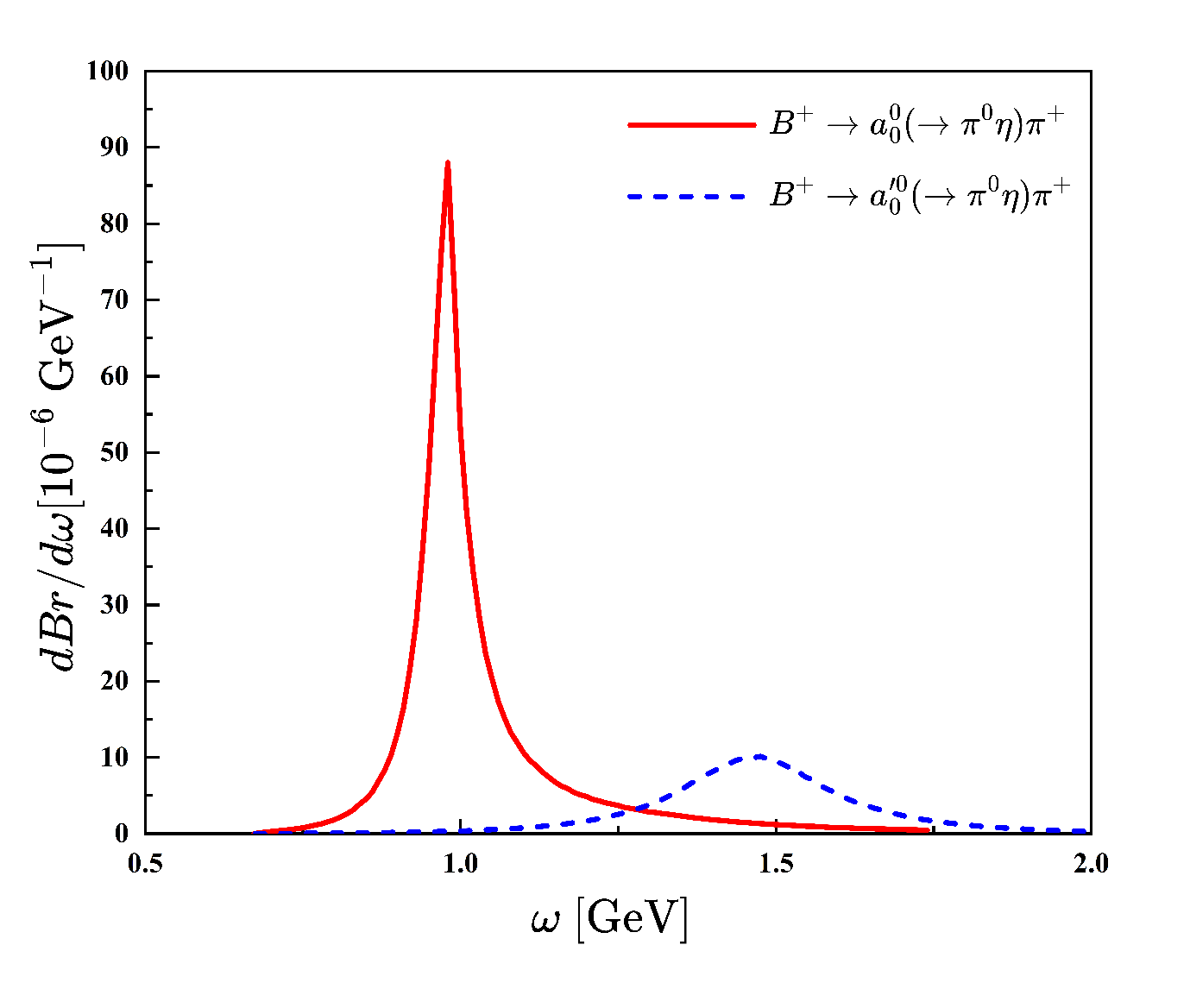}\quad}
	\subfigure[]{\includegraphics[width=0.3\textwidth]{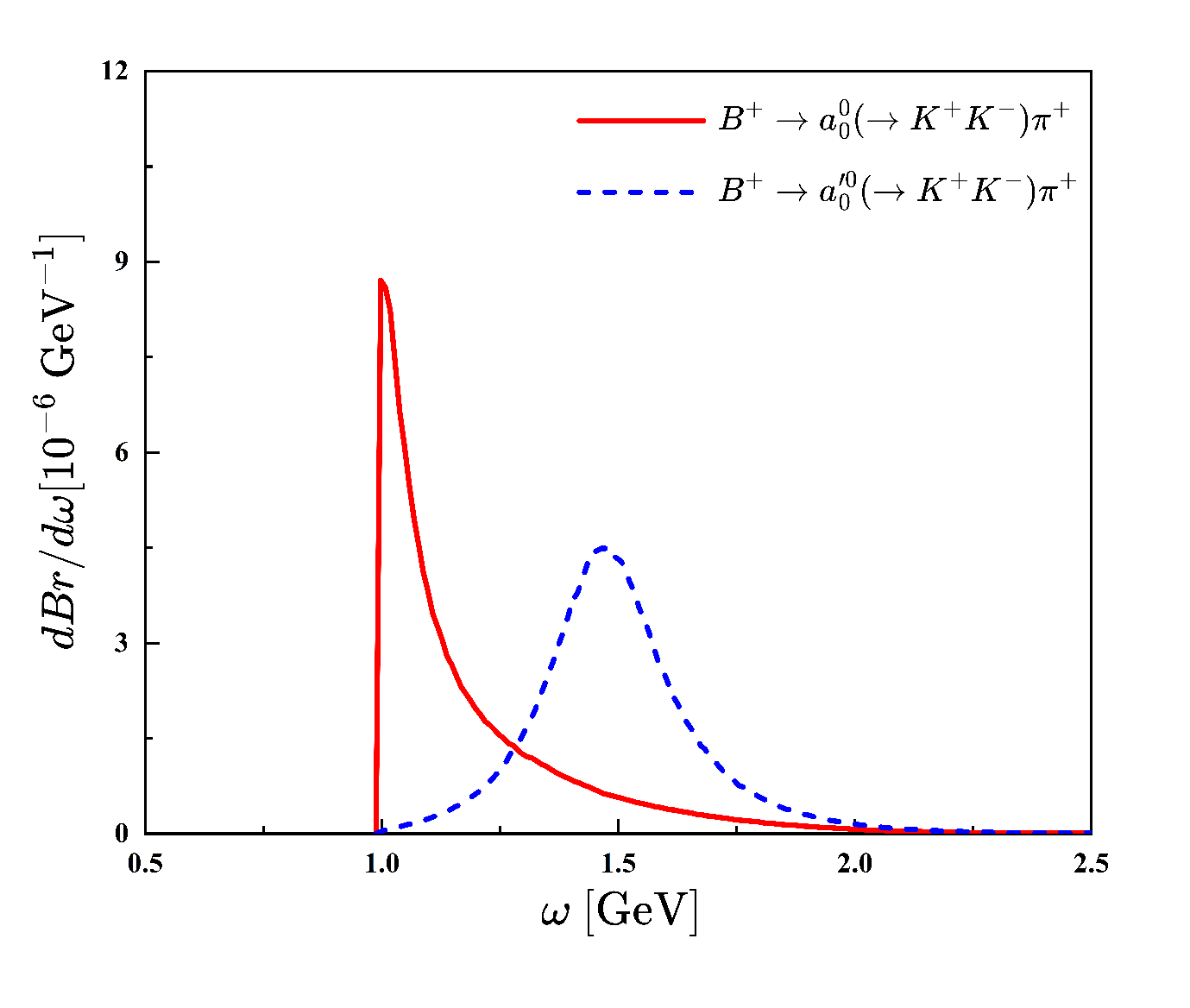}\quad}
	\caption{The dependencies of the differential branching ratios of the quasi-two-body decays $B\to a^{(\prime)}_0(\to \pi\eta, K\bar K)h$ on the invariant mass $\omega$.}\label{distri}
\end{figure}

By comparing the $B^+\to a^0_0(\to\pi^0\eta)\pi^+$ and  $B^+\to a^0_0(\to K^+K^-)\pi^+$ decay spectra shown in Fig.~\ref{distri}(a), one can find that the  branching ratio of the decay $B^+\to a^0_0(\to K^+K^-)\pi^+$ is much smaller than that of the decay $B^+\to a^0_0(\to \pi^0\eta)\pi^+$, as expected. It is because that the $a_0$ contribution from $K\bar K$ mode is highly suppressed from phase space. There exists the similar situation between the decays $B^+\to a^0_0(\to \pi^0\eta)K^+$ and  $B^+\to a^0_0(\to K^+K^-)K^+$, shown in  Fig.~\ref{distri}(b). Since the branching ratios of the strong decays $a_0^\prime \to \pi\eta$ and $a_0^\prime \to K\bar K$ are comparable, as mentioned above, the curves of the differential distributions for the branching ratios of the decays $B^+\to a^{\prime+}_0(\to\pi^+\eta) K^0$ and $B^+\to a^{\prime+}_0(\to K^+\bar K^0) K^0$ are close to each other, which are shown in Fig.~\ref{distri}(c). While because of the isospin relation $\mathcal{B}r(a_0^{\prime0} \to K^+K^-)=\mathcal{B}r(a^{\prime0}_0\to K\bar K)/2\approx \mathcal{B}r(a_0^{\prime0} \to \pi^0\eta)/2$, the peak for the  $B^+\to a^{\prime0}_0(\to K^+K^-)\pi^+$ decay spectrum becomes much lower than that 
for the $B^+\to a^{\prime0}_0(\to \pi^0\eta)\pi^+$ decay spectrum, which can be found in Fig.~\ref{distri} (d). From Fig.~\ref{distri} (e), one can find that the decay modes $B^+\to a^{(\prime)0}_0(\to\pi^0\eta)\pi^+$ are mainly mediated by the resonance $a_0$, and the contribution from the resonance $a_0^\prime$ is very small. This is because that the channel $a_0\to \pi^0\eta$ is absolutely dominant in all the $a_0$ decays. While if replaced $\pi^0\eta$ with $K^+K^-$ in these two decays, the situation will be different. Comparing the $B^+\to a^0_0(\to K^+K^-)\pi^+$ and $B^+\to a^{\prime0}_0(\to K^+K^-)\pi^+$ decay spectra shown in Fig.~\ref{distri}(f), it is obvious that the difference between the corresponding branching ratios should not be very large.

\begin{table}[h!]
\caption{The CP violations (in units of $\%$) of the decays $B\to a_0(\to KK,\pi\eta)h$. }
\begin{center}
\begin{tabular}{@{\extracolsep{0.5cm}}ccccccccc}
\hline
Decay modes&This work&Ref. \cite{chaij}& Refs. \cite{shenyl,zhangzhiqing}& Ref. \cite{zhenghaiyang}\\
\hline
$B^{+}\to a_0^+(\rightarrow\pi^+\eta) K^0$ &$5.02^{+1.12 +0.00}_{-1.72 -1.53}$&$3.72^{+5.64}_{-4.46}$&\multirow{2}{*}{4}&\multirow{2}{*}{$0.63^{+54.28}_{-2.96}$}\\
$B^{+}\to a_0^+(\rightarrow K^+ \bar K^0)K^0$ &$3.88^{+0.53 +1.06}_{-0.68 -1.64}$&$6.10^{+7.71}_{-7.90}$&  &\\
\hline
$B^{+}\to a_0^0(\rightarrow\pi^0\eta)K^+$  &$-23.7^{+22.2 +26.9}_{-21.1 -17.1}$&$-21.3^{+8.4}_{-10.6}$&\multirow{2}{*}{-70}&\multirow{2}{*}{$-13.3^{+57.3}_{-40.7}$}\\
$B^{+}\to a_0^0(\rightarrow K^+K^-)K^+$ &$-47.2^{+10.2 +5.2}_{-8.8-0.0}$ &$-26.4^{+6.9}_{-7.9}$ &  &\\
\hline
$ B^0\to a_0^0(\rightarrow\pi^0\eta) K^0$ &$-47.5^{+2.5 +27.5}_{-5.5 -0.0}$&$-43.2^{+6.0}_{-11.8}$&\multirow{2}{*}{-17}&\multirow{2}{*}{$-7.6^{+68.3}_{-30.9}$}\\
$ B^0\to a_0^0(\rightarrow K^+K^-)K^0$ &$-20.3^{+4.3 +14.9}_{-2.7 -14.7}$ &$-27.5^{+8.9}_{-2.7}$&  &\\
\hline
$B^0\to a_0^-(\rightarrow\pi^-\eta)K^+$ &$-49.1^{+24.1 +11.1}_{-17.9 -0.0}$ &$-83.2^{+4.2}_{-13.8}$ &\multirow{2}{*}{-70}&\multirow{2}{*}{$-19.1^{+76.1}_{-42.6}$} \\
$B^0\to a_0^-(\rightarrow K^-K^0)K^+$ &$-50.9^{+8.9 +9.9}_{-7.1 -7.1}$ &$-69.7^{+2.1}_{-4.9}$ &  &\\
\hline
$B^{+}\to a_0^+(\rightarrow\pi^+\eta)\pi^0$&$8.93^{+4.47 +3.37}_{-13.20 -16.00}$ &$56.3^{+3.1}_{-7.9}$ &\multirow{2}{*}{{$70\sim 80$}}&\multirow{2}{*}{$-32.8^{+21.3}_{-22.6}$} \\
$B^{+}\to a_0^+(\rightarrow K^+\bar K^0)\pi^0$&$13.3^{+5.9 +2.8}_{-8.0 -10.2}$&$38.2^{+5.0}_{-7.8}$ &  & \\
\hline
$B^{+}\to a_0^0(\rightarrow\pi^0\eta)\pi^+$&$-0.90^{+0.72 +1.11}_{-0.72 -1.46}$ &$26.5^{+5.4}_{-6.7}$ &\multirow{2}{*}{14}&\multirow{2}{*}{$-8.8^{+1.1}_{-1.0}$} \\
$B^{+}\to a_0^0(\rightarrow K^+K^-)\pi^+$&$-0.01^{+0.51 +0.80}_{-0.47 -1.16}$ &$24.1^{+7.0}_{-7.0}$ &  & \\
\hline
$B^0\to a_0^0(\rightarrow\pi^0\eta)\pi^0$&$-51.9^{+2.9 +0.9}_{-7.1 -9.1}$ &$84.1^{+7.5}_{-8.1}$ &\multirow{2}{*}{{$70\sim 80$}}&\multirow{2}{*}{$-30.9^{+30.9}_{-25.4}$}\\
$B^0\to a_0^0(\rightarrow K^+K^-)\pi^0$&$-47.2^{+0.0 +1.2}_{-5.8 -9.8}$ &$79.4^{+7.9}_{-11.6}$ &  & \\
\hline
$B^0\to a_0^-(\rightarrow\pi^-\eta)\pi^+$&$1.31^{+1.49 +0.53}_{-1.55 -0.69}$&$20.6^{+3.8}_{-4.5}$&\multirow{2}{*}{--} &\multirow{2}{*}{$1.1^{+16.5}_{-16.9}$} \\
$B^0\to a_0^-(\rightarrow K^-K^0)\pi^+$& $4.94^{+0.96 +0.02}_{-0.93 -0.14}$&$17.8^{+3.9}_{-4.1}$ & & \\
\hline
$ B^0\to a_0^+(\rightarrow\pi^+\eta)\pi^-$&$85.5^{+6.5 +1.5}_{-8.5 -0.5}$&$68.3^{+7.3}_{-9.5}$ &\multirow{2}{*}{--}&\multirow{2}{*}{$-5.0^{+16.6}_{-15.5}$} \\
$B^0\to a_0^+(\rightarrow K^+\bar K^0)\pi^-$&$84.0^{+4.0 +2.0}_{-7.0 -2.0}$&$70.5^{+6.9}_{-8.0}$ & & \\
\hline
\end{tabular}
\end{center}
\end{table}

Now we turn to the evaluations of the CP-violating asymmetries for the decays we are considering. Thus
far the CP violation has not been observed in any B decays involving a scalar meson. The direct CP violation is defined as
\begin{equation}
A_{CP}=\frac{\bar A-A}{\bar A+A}= 
	\begin{cases}
	\frac{2z\sin\gamma\sin\delta}{1-2z\cos\gamma\cos\delta+z^2} \qquad \text{for $b\to s$ transition,}\\[0.15cm]
	\frac{2z\sin\alpha\sin\delta}{1+2z\cos\alpha\cos\delta+z^2} \qquad \text{for $b\to d$ transition,}
	\end{cases}
\end{equation}
where $z=\left|\frac{V_{tb}V^*_{tq}}{V_{ub}V^*_{uq}}\frac{P}{T}\right|$ with $q=d,s$, and $\delta$ is the relative strong phase between the tree ($T$) and the penguin ($P$) amplitudes. From upper formula, one can find that the direct CP violation depends
on the ratio of the tree and penguin amplitudes except for the CKM elements and the strong phases.  When the ratio is too large or too small, the direct CP asymmetry will be very small. For example, for the decays $B^+\to a^{(\prime)+}_0K^0$, the penguin operators give the dominant contribution, and the tree operators suffer from the CKM suppression, so their direct CP asymmetries are small and only a few percent even smaller, which have been proved in both the PQCD and the QCDF approaches, no matter in two-body or three-body framework. Another two examples are the decays $B^+\to a^0_0\pi^+$ and  $B^0\to a^-_0\pi^+$, whose direct CP violations given both in the QCDF and our results are small. Certainly, it is similar when replacing $a_0$ with $a_0^\prime$ in these decays. While the comparable tree and penguin amplitudes implying strong inferences between them will induce large direct CP asymmetries. For example, the direct CP violations
of the decays $B^0\to a^-_0 K^+, a^+_0\pi^-, a^0_0\pi^0$ predicted by the PQCD approach under three-body framework can (almost) reach up to $50\%$ even more in magnitude. Such large values were also supported  by the previous PQCD calculations \cite{shenyl,chaij,zhangzhiqing}. While the results obtained in the QCDF approach are relatively small but with large uncertainties. Furthermore, for some of our considered decays, the signs of the direct CP violations are just contrary between the PQCD and the QCDF predictions, which indicate that there exists obviously a difference in the sources of the strong phases between these two approaches. Under the PQCD approach, the annihilation diagrams are not sufficient power suppressed in some cases. Usually a large imaginary part from these annihilation diagram amplitudes will appear, which provides an important source of the strong phases in the PQCD approach, while it is missing for the QCDF approach. Certainly, there also exists significant difference for the direct CP violations between these PQCD calculations, which is mainly induced by the discrepancies from the LCDAs describing the resonance $a^{(\prime)}_0$. Further clarifying these discrepancies are helpful to probe the inner structures of these two resonances.  

\begin{table}[h!]
\caption{The CP violations (in units of $\%$) of the decays $B\to a_0^{\prime0}(\to KK,\pi\eta)h$.}
\begin{center}
\begin{tabular}{@{\extracolsep{1cm}}ccccccccc}
\hline
CP violation&This work&Ref. \cite{chaij}&Ref. \cite{zhenghaiyang}\\
\hline
$B^{+}\to a_0^{\prime+}(\rightarrow\pi^+\eta)K^0$ &$2.36^{+0.46 +1.06}_{-0.54 -0.77}$&$0.30^{+0.82}_{-0.94} $&\multirow{2}{*}{$0.22^{+0.59}_{-47.94}$}\\
$B^{+}\to a_0^{\prime+}(\rightarrow K^+\bar K^0)K^0$ &$2.69^{+0.46 +1.11}_{-0.49 -0.82}$&$0.50^{+0.61}_{-0.61}$ & \\
\hline
$B^{+}\to a_0^{\prime0}(\rightarrow\pi^0\eta)K^+$ &$-42.2^{+4.2 +14.3}_{-2.5 -13.7}$ &$-23.6^{+4.8}_{-2.7}$ &\multirow{2}{*}{$-1.79^{+16.63}_{-17.77}$}  \\
$B^{+}\to a_0^{\prime0}(\rightarrow K^+K^-)K^+$ &$-41.9^{+3.1 +14.1}_{-2.6 -14.4}$ &$-22.7^{+2.7}_{-3.3}$ &  \\
\hline
$B^0\to a_0^{\prime0}(\rightarrow\pi^0\eta)K^0$ &$-10.9^{+4.4 +6.3}_{-3.1 -9.3}$&$-7.50^{+1.30}_{-3.32}$&\multirow{2}{*}{$-0.74^{+8.20}_{-20.70}$}\\
$B^0\to a_0^{\prime0}(\rightarrow K^+K^-)K^0$ &$-10.6^{+4.3 +6.0}_{-3.2 -9.1}$ &$-6.30^{+0.81}_{-3.98}$ &\\
\hline
$B^0\to a_0^{\prime-}(\rightarrow \pi^-\eta)K^+$ &$-41.8^{+3.1 +13.4}_{-2.9 -14.0}$ &$-46.0^{+5.2}_{-4.4} $&\multirow{2}{*}{$-0.65^{+11.32}_{-23.81}$}\\
$B^0\to a_0^{\prime-}(\rightarrow K^-K^0)K^+$ &$-41.4^{+2.9 +13.7}_{-2.6 -14.3}$ &$-46.7^{+4.4}_{-3.6}$&\\
\hline
$B^{+}\to a_0^{\prime+}(\rightarrow\pi^+\eta)\pi^0$&$16.9^{+10.0 +2.5}_{-0.7-0.1}$&$-15.2^{+5.8}_{-7.5} $&\multirow{2}{*}{$-24.6^{+25.6}_{-23.9}$}\\
$B^{+}\to a_0^{\prime+}(\rightarrow K^+\bar K^0)\pi^0$&$18.0^{+9.3 +2.3}_{-0.3 -0.0}$&$-19.2^{+7.1}_{-9.2} $&\\
\hline
$B^{+}\to a_0^{\prime0}(\rightarrow\pi^0\eta)\pi^+$& $0.66^{+0.32 +0.31}_{-0.25 -0.80}$&$1.00^{+3.49}_{-4.52}$&\multirow{2}{*}{$-13.1^{+5.0}_{-4.4}$}\\
$B^{+}\to a_0^{\prime0}(\rightarrow K^+K^-)\pi^+$&$0.81^{+0.31 +0.27}_{-0.25 -0.75}$&$-0.10^{+3.94}_{-2.50}$& \\
\hline
$B^0\to a_0^{\prime0}(\rightarrow\pi^0\eta)\pi^0$&$-42.5^{+0.0 +1.8}_{-4.2 -10.7}$&$24.3^{+18.4}_{-13.9} $&\multirow{2}{*}{$-13.8^{+35.9}_{-20.4}$} \\
$B^0\to a_0^{\prime0}(\rightarrow K^+K^-)\pi^0$&$-42.2^{+0.0 +1.7}_{-4.4 -10.8}$ &$26.1^{+4.8}_{-10.9} $& \\
\hline
$B^0\to a_0^{\prime-}(\rightarrow \pi^-\eta)\pi^+$&$7.78^{+0.68 +0.55}_{-0.62 -0.85}$&$26.1^{+4.8}_{-4.6} $&\multirow{2}{*}{$-0.21^{+82.80}_{-83.60}$}\\
$B^0\to a_0^{\prime-}(\rightarrow K^-K^0)\pi^+$&$7.99^{+0.64 +0.59}_{-0.60 -0.85}$&$25.8^{+5.2}_{-4.7}$& \\
\hline
$B^0\to a_0^{\prime+}(\rightarrow \pi^+\eta)\pi^-$&$82.3^{+3.0 +2.8}_{-5.7 -1.8}$&$28.5^{+6.7}_{-4.6} $&\multirow{2}{*}{$-5.2^{+62.1}_{-67.0}$} \\
$B^0\to a_0^{\prime+}(\rightarrow K^+\bar K^0)\pi^-$&$82.6^{+2.5 +1.8}_{-6.3 -2.4}$&$24.8^{+6.5}_{-6.8} $& \\
\hline
\end{tabular}
\end{center}
\end{table}
\section{Conclusion}
\label{summary}
In this work, we have discussed contributions from the isovector scalar resonances $a^{(\prime)}_0$ to the quasi-two-body decays $B^0\to a^{(\prime)}_0h\to \pi\eta(K\bar K)h$ with $h$ being a pion or a kaon, based on the PQCD approach. Different from the previous two-body framework, the dimeson DAs were introduced to describe the strong dynamics between the S-wave resonances $a^{(\prime)}_0$ and the $\pi\eta(K\bar K)$ pair, where the Gegenbauer coefficient was determined from the data. Adopting experimental inputs for the time-like form factors involved in the dimeson DAs, we have calculated the branching ratios and the direct CP asymmetries of the decays $B^0\to a^{(\prime)}_0h\to \pi\eta(K\bar K)h$. It is found that most of the branching ratios of the two-body decays $B\to a^{(\prime)}_0(\pi, K)$ obtained with the narrow width approximation are closer to the QCDF results compared to the previous PQCD calculations, under both the three-body and two-body frameworks. For the decays $B^0\to a^{(\prime)-}_0\pi^+$ and $B^+\to a^{(\prime)0}_0\pi^+$, we obtained unexpectedly large branching ratios by assuming $a^{(\prime)}_0$ as a simple $q\bar q$ state, which are urgent to be examined in experiments. For the direct CP violations, there exist obvious differences among different calculations which are mainly induced by the large uncertainties from the strong phases for the decays we are considering. 

\section*{Acknowledgment}
This work is partly supported by the National Natural Science
Foundation of China under Grant No. 11347030, by the Program of
Science and Technology Innovation Talents in Universities of Henan
Province 14HASTIT037, and the Natural Science Foundation of Henan
Province under grant no. 232300420116.

\appendix
\section{Some relevant functions}
The explicit expressions of the hard functions $h_i$ with $i=a, \cdots, h$ are obtained from the Fourier transform of the hard kernels and given as
\begin{equation}
\begin{aligned}
h_i\left(\alpha, \beta, b_1, b_2\right) & =h_1\left(\alpha, b_2\right) \times h_2\left(\beta, b_1, b_2\right), \\
h_1\left(\alpha, b_2\right) & = \begin{cases}K_0\left(\sqrt{\alpha} b_2\right), & \alpha>0 ;\\
K_0\left(i \sqrt{-\alpha} b_2\right), & \alpha<0;\end{cases} \\
h_2\left(\beta, b_1, b_2\right) & = \begin{cases}\theta\left(b_2-b_1\right) I_0\left(\sqrt{\beta} b_1\right) K_0\left(\sqrt{\beta} b_2\right)+\left(b_1 \leftrightarrow b_2\right), & \beta>0 ; \\
\theta\left(b_2-b_1\right) I_0\left(\sqrt{-\beta} b_1\right) K_0\left(i \sqrt{-\alpha} b_2\right)+\left(b_1 \leftrightarrow b_2\right), & \beta<0 ;\end{cases}
\end{aligned}
\end{equation}
where $K_0$ and $I_0$ are modified Bessel functions with $K(ix)=\frac{\pi}{2}(-N_0(x)+iJ_0(x))$ and $J_0$ is Bessel function.
The hard scales $t_i$ are chosen as the maximum of the virtuality of the internal momentum transition in the hard amplitudes and listed as following
\begin{equation}
\begin{aligned}
t_a & =\max \left\{m_{B} \sqrt{\left|\alpha_e\right|}, m_{B} \sqrt{\left|\beta_a\right|}, 1 / b_3, 1 / b_1\right\}, && t_b=\max \left\{m_{B} \sqrt{\left|\alpha_e\right|}, m_{B} \sqrt{\left|\beta_b\right|}, 1 / b_1, 1 / b_3\right\}, \\
t_c & =\max \left\{m_{B} \sqrt{\left|\alpha_e\right|}, m_{B} \sqrt{\left|\beta_c\right|}, 1 / b_1, 1 / b\right\}, && t_d=\max \left\{m_{B} \sqrt{\left|\alpha_e\right|}, m_{B} \sqrt{\left|\beta_d\right|}, 1 / b_1, 1 / b\right\}, \\
t_e & =\max \left\{m_{B}\sqrt{\left|\alpha_a\right|}, m_{B} \sqrt{\left|\beta_e\right|}, 1 / b_3, 1 / b_1\right\}, && t_f=\max \left\{m_{B} \sqrt{\left|\alpha_a\right|}, m_{B} \sqrt{\left|\beta_f\right|}, 1 / b_3, 1 / b_1\right\},\\
t_g & =\max \left\{m_{B} \sqrt{\left|\alpha_a\right|}, m_B \sqrt{\left|\beta_g\right|}, 1 / b_3, 1 / b\right\}, && t_h=\max \left\{m_{B} \sqrt{\left|\alpha_a\right|}, m_{B} \sqrt{\left|\beta_h\right|}, 1 / b, 1 / b_3\right\},
\end{aligned}
\label{scale}
\end{equation}
where
\begin{equation}
\begin{aligned}
& \beta_a=r_b^2-1+z, && \beta_b=x_B-\eta,  \\
& \beta_c=(x_B-(1-\eta)(1-x_3))z, &&\beta_d=(x_B-(1-\eta)x_3)z, \\
&\beta_e=r_b^2+(x_B-(1-\eta)(1-x_3))z,&& \beta_f=(x_B-\eta-(1-\eta)x_3)(1-z),  \\
& \beta_g=z-1, && \beta_h=-(\eta+(1-\eta)x_3),\\
&\alpha_e=x_Bz , &&\alpha_a=-(\eta+(1-\eta) x_3)(1-z).\\
&
\end{aligned}
\end{equation}
The Sudakov factors can be written as
\begin{equation}
\begin{aligned}
S_{a b}(t)= & s\left(\frac{m_{B}}{\sqrt{2}} x_B, b_B\right)+s\left(\frac{m_{B}}{\sqrt{2}} x_3, b_3\right)+\frac{5}{3} \int_{1 / b_B}^t \frac{d \mu}{\mu} \gamma_q(\mu)+2 \int_{1 / b_3}^t \frac{d \mu}{\mu} \gamma_q(\mu), \\
S_{c d}(t)= & s\left(\frac{m_{B}}{\sqrt{2}} x_B, b_B\right)+s\left(\frac{m_{B}}{\sqrt{2}} z, b\right)+s\left(\frac{m_{B}}{\sqrt{2}} (1-z), b\right)+s\left(\frac{m_{B}}{\sqrt{2}} x_3, b_B\right) \\
& +\frac{11}{3} \int_{1 / b_B}^t \frac{d \mu}{\mu} \gamma_q(\mu)+2 \int_{1 / b}^t \frac{d \mu}{\mu} \gamma_q(\mu), \\
S_{ef}(t)= & s\left(\frac{m_{B}}{\sqrt{2}} x_B, b_1\right)+s\left(\frac{m_{B}}{\sqrt{2}} z, b_3\right)+s\left(\frac{m_{B}}{\sqrt{2}} (1-z), b_3\right)+s\left(\frac{m_{B}}{\sqrt{2}} x_3, b_3\right) \\
& +\frac{5}{3} \int_{1 / b_1}^t \frac{d \mu}{\mu} \gamma_q(\mu)+4 \int_{1 / b_3}^t \frac{d \mu}{\mu} \gamma_q(\mu),\\
S_{gh}(t)= & s\left(\frac{m_{B}}{\sqrt{2}} z, b\right)+s\left(\frac{m_{B}}{\sqrt{2}} (1-z), b\right)+s\left(\frac{m_{B}}{\sqrt{2}} x_3, b_3\right) \\
& +2 \int_{1 / b}^t \frac{d \mu}{\mu} \gamma_q(\mu)+2 \int_{1 / b_3}^t \frac{d \mu}{\mu} \gamma_q(\mu),
\end{aligned}
\label{suda}
\end{equation}
where the quark anomalous dimension $\gamma_q(\mu)=-\alpha_s(\mu)/\pi$ and the definition of the function $s(q,b)$ can be found in \cite{zhaoyc}.
As we know, the double logarithms $\alpha_sln^2x$ produced by the radiative corrections are not
small expansion parameters when the end point region is important, in order to improve the
perturbative expansion, the threshold resummation of these logarithms to all order is needed, which
leads to a quark jet function
\be
S_t(x)=\frac{2^{1+2c}\Gamma(3/2+c)}{\sqrt{\pi}\Gamma(1+c)}[x(1-x)]^c,
\en
with $c=0.3$. It is effective to smear the end point singularity with a momentum fraction $x\to0$.
\section{Factorization formulae}
If the $a_0$ is emitted with $B$ transitting to $\pi$, the factorization formulae for the factorizable emission diagrams are given as
\be
\mathcal{F}^{\prime LL}_e&=&8\pi
C_FM^4_{B} F_{s}(s)/\mu_s\int_0^1 dx_{B} dx_3\, \int_{0}^{\infty} b_B db_B
b_3 db_3\, \phi_{B}(x_B,b_B)\non && \left\{\left[(1-\eta)((\eta-1)x_3-1)\phi^A_\pi-r_\pi((1+\eta+2(\eta-1)x_3)\phi^P_\pi+(\eta-1)(2x_3-1)\phi^T_\pi)\right]\right.\non &&\left.
\times \alpha_s(t^\prime_a)S_t(z)h(\alpha^\prime_e,\beta^\prime_a,b_B,b_3)\exp[-S_{ab}(t^\prime_a)]
+[\eta(\eta-1)x_B\phi^A_\pi\right.\non &&\left.+2r_\pi(\eta-1+\eta x_B)\phi^P_\pi] \alpha_s(t^\prime_b)S_t(x_B) h_e(\alpha^\prime_e,\beta^\prime_b,b_3,b_B)\exp[-S_{ab}(t^\prime_b)]\right\},\\	
\mathcal{F}^{LR}_e &=&-\mathcal{F}^{LL}_e,			\\
\mathcal{F}^{\prime SP}_e&=& 16\pi C_FM^4_{B}F_{s}(s)/\mu_s\int_0^1 dx_{B} dx_3\, \int_{0}^{\infty} b_B db_Bb_3 db_3\phi_{B}(x_B,b_B)\left\{\left[(\eta-1)\phi^A_\pi\right.\right.\non && \left.\left.
+r_\pi(((\eta-1)x_3-2)\phi^P_\pi-(\eta-1)x_3\phi^T_\pi) \right]\alpha_s(t^\prime_a)S_t(z)h(\alpha^\prime_e,\beta^\prime_a,b_B,b_3)\right.\non && \left.
\times\exp[-S_{ab}(t^\prime_a)]
+\alpha_s(t^\prime_b)S_t(x_B)h(\alpha^\prime_e,\beta^\prime_a,b_3,b_B)\exp[-S_{ab}(t^\prime_b)]\right.\non&&\left.\times[x_B(\eta-1)\phi_\pi^A+2r_\pi(x_B+\eta-1)\phi^P_\pi]\right\}.
\en
The amplitudes for the nonfactorizable emission diagrams are written as
\be
\mathcal{M}^{\prime LL}_{e}&=&32\pi C_F m_{B}^4/\sqrt{2N_C}\int_0^1 d x_{B} dz dx_{3}
\int_{0}^{\infty} b_B db_B b db\,\phi_{B}(x_B,b_B)\phi^0\non &&
\times\left\{\left[(\eta^2-1)(x_B+z-1)\phi^A_\pi+r_\pi(\eta(x_B+z)(\phi^P_\pi+\phi^T_\pi)+(\eta-1)x_3\right.\right.\non && \left.\left.\times(\phi^P_\pi-\phi^T_\pi)-2\eta\phi^P_\pi)\right] \alpha_s(t^\prime_{c})h(\beta^\prime_c,\alpha^\prime_e,b_B,b)\exp[-S_{cd}(t^\prime_c)]\right.\non && \left.+\left[(\eta-1)((1-\eta)x_3-x_1+z)\phi^A_\pi-\eta r_\pi(x_1-z)(\phi^P_\pi-\phi^T_\pi)\right.\right.\non && \left.\left.-r_\pi(\eta-1)x_3(\phi^P_\pi+\phi^T_\pi)\right]\alpha_s(t^\prime_{d})h(\beta^\prime_d,\alpha^\prime_e,b,b_B)\exp[-S_{cd}(t^\prime_d)]\right\}, \label{nfe1}\\
\mathcal{M}^{\prime LR}_{e}&=&32\pi C_F m_{B}^4\sqrt\eta/\sqrt{2N_C}\int_0^1 d x_{B} dz dx_{3}\int_{0}^{\infty} b_B db_B b db\,\phi_{B}(x_B,b_B)\non &&
\times\left\{\left[(\eta-1)(x_B+z-1)\phi^A_\pi(\phi^s+\phi^t)+r_\pi(\eta(1-x_3)+x_3)(\phi^P_\pi+\phi^T_\pi)(\phi^s-\phi^t)\right.\right.\non &&\left.\left.-(x_B+z-1)(\phi^P_\pi-\phi^T_\pi)(\phi^s+\phi^t)\right] \alpha_s(t^\prime_{c})h(\beta^\prime_c,\alpha^\prime_e,b_B,b)\exp[-S_{cd}(t^\prime_c)]
\right.\non && \left.+\left[(x_B-z)(\phi^s-\phi^t)((1-\eta)\phi^A_\pi+r_\pi(\phi^P_\pi-\phi^T_\pi))+(\eta-1)r_\pi x_3(\phi^s+\phi^t)\right.\right.\non &&\left.\left.\times(\phi^P_\pi+\phi^T_\pi)\right]\alpha_s(t^\prime_{d})h(\beta^\prime_d,\alpha^\prime_e,b,b_B)\exp[-S_{cd}(t^\prime_d)]\right\},\\
\mathcal{M}^{\prime SP}_{e}&=&32\pi C_f m_{B}^4/\sqrt{2N_C}\int_0^1 d x_{B} dz dx_{3}
\int_{0}^{\infty} b_B db_B b db\,\phi_{B}(x_B,b_B)\phi^0\non &&
\times\left\{[(\eta-1)(x_B+z+(\eta-1)x_3-\eta-1)\phi^A_\pi+r_\pi(\eta(x_B+z)(\phi^P_\pi-\phi^T_\pi)\right.\non &&  \left.+(\eta-1)x_3(\phi^P_\pi+\phi^T_\pi))-2\eta\phi^P_\pi] \alpha_s(t^\prime_{c})h(\beta^\prime_c,\alpha^\prime_e,b_B,b)\exp[-S_{cd}(t^\prime_c)]\right.\non && \left.-[(x_B-z)((\eta^2-1)\phi^A_\pi+\eta r_\pi(\phi^P_\pi+\phi^T_\pi))+(\eta-1)r_\pi x_3(\phi^P_\pi-\phi^T_\pi)] \right.\non && \left.\times
\alpha_s(t^\prime_{d})h(\beta^\prime_d,\alpha^\prime_e,b,b_B)\exp[-S_{cd}(t^\prime_d)]\right\}.
\en
In the annihilation diagrams, if $a_0$ is the upper meson, that is in the $b$ quark side, the nonfactorizable annihilation amplitudes are given as 
\be
\mathcal{M}^{\prime LL}_{a}&=& 32\pi C_f m_{B}^4/\sqrt{2N_C}\int_0^1 d x_{B} dz dx_{3} \int_{0}^{\infty} b_B db_B b db\,\phi_{B}(x_B,b_B)\non && 
\times\left\{\left[((x_B+z-1)\eta^2+\eta-x_B-z)\phi^0\phi^A_\pi+\sqrt\eta r_\pi((1-x_B-z)(\phi^s+\phi^t)\right.\right.\non && \left.\left. \times(\phi^P_\pi-\phi^T_\pi)+(x_3-\eta(x_3-1))(\phi^s-\phi^t)(\phi^P_\pi+\phi^T_\pi)-4\phi^s\phi^P_\pi)
\right]\right.\non && \left.\times\alpha_s(t^\prime_{e})h(\beta^\prime_e,\alpha^\prime_a,b_B,b)\exp[-S_{ef}(t^\prime_e)]\right.\non &&\left. -\left[(\eta-1)(\eta(z+x_3-x_B-1)-x_3+1)\phi^0\phi^A_\pi+\sqrt\eta r_\pi((\eta-1)(1-x_3)\right.\right.\non &&\left.\left.\times (\phi^s+\phi^t)(\phi^P_\pi-\phi^T_\pi)+(x_B-z)(\phi^s-\phi^t)(\phi^P_\pi+\phi^T_\pi))
\right]\right.\non &&\left.\times \alpha_s(t^\prime_{f})h(\beta^\prime_f,\alpha^\prime_a,b,b_B)\exp[-S_{ef}(t^\prime_f)]\right\},\\
\mathcal{M}^{\prime LR}_{a}&=& 32\pi C_F m_{B}^4/\sqrt{2N_C}\int_0^1 d x_{B} dz dx_{3} \int_{0}^{\infty} b_B db_B b db\,\phi_{B}(x_B,b_B)\non && 
\times\left\{\left[(\eta-1)(r_\pi(1+x_3)\phi^0(\phi^P_\pi-\phi^T_\pi)+\sqrt\eta(x_B+z-2)\phi^A_\pi(\phi^s+\phi^t))\right.\right.\non && \left.\left.+\eta r_\pi \phi^0((x_B+z)(\phi^P_\pi+\phi^T_\pi)-4\phi^P_\pi)\right]\alpha_s(t^\prime_{e})h(\beta^\prime_e,\alpha^\prime_a,b_B,b)\exp[-S_{ef}(t^\prime_e)]\right.\non &&\left.+\left[\sqrt\eta(\eta-1)(x_B-z)\phi^A_\pi(\phi^s+\phi^t)+r_\pi \phi^0(\eta(x_B-z)(\phi^P_\pi+\phi^T_\pi)\right.\right.\non && \left.\left.+(\eta-1)(1-x_3)(\phi^P_\pi-\phi^T_\pi))\right]\alpha_s(t^\prime_{f})h(\beta^\prime_f,\alpha^\prime_a,b,b_B)\exp[-S_{ef}(t^\prime_f)]\right\},\\
\mathcal{M}^{\prime SP}_{a}&=& 32\pi C_f m_{B}^4/\sqrt{2N_C}\int_0^1 d x_{B} dz dx_{3} \int_{0}^{\infty} b_B db_B b db\,\phi_{B}(x_B,b_B)\left\{\left[(1-\eta) \right.\right.\non && \left.\left. 
\times(\eta(x_B+z+x_3-2)-x_3+1)\phi^0\phi^A_\pi+\sqrt\eta r_\pi((x_B+z-1)(\phi^s-\phi^t)\right.\right.\non &&\left.\left.\times(\phi^P_\pi+\phi^T_\pi)+((\eta-1)x_3-\eta)(\phi^s+\phi^t)(\phi^P_\pi-\phi^T_\pi)+4\phi^s\phi^P_\pi)\right] \right.\non &&\left.\times \alpha_s(t^\prime_{e})h(\beta^\prime_e,\alpha^\prime_a,b_B,b)\exp[-S_{ef}(t^\prime_e)]+\left[(1-\eta^2)(x_B-z)\phi^0\phi^A_\pi+\sqrt\eta\right.\right.\non && \left.\left. \times ((x_B-z)(\phi^s+\phi^t)(\phi^P_\pi-\phi^T_\pi)+(1-\eta)
(x_3-1)(\phi^s-\phi^t)(\phi^P_\pi+\phi^T_\pi))
\right]\right.\non && \left. \times \alpha_s(t^\prime_{f})h(\beta^\prime_f,\alpha^\prime_a,b,b_B)\exp[-S_{ef}(t^\prime_f)]\right\}.
\en
The factorizable annihilation amplitudes are listed as
\be
\mathcal{F}^{\prime LL}_{a}&=& 8\pi C_F m_{B}^4f_B/\sqrt{2N_C}\int_0^1  dz dx_{3} \int_{0}^{\infty} b db b_3 db_3\non && 
\times\left\{[(\eta-1)(1+(\eta-1)x_3)\phi^0\phi^A_\pi-2\sqrt\eta r_\pi \phi^s((\eta-1)x_3(\phi^P_\pi-\phi^T_\pi)+2\phi^P_\pi)]\right.\non &&\left.
\times \alpha_s(t^\prime_{g})S_t(z)h(\alpha^\prime_a,\beta^\prime_g,b_3,b)\exp[-S_{gh}(t^\prime_g)]
\right.\non &&\left.-[z(\eta-1)\phi^0\phi^A_\pi+2\sqrt\eta r_\pi \phi^P_\pi((\eta-1)(\phi^s-\phi^t)-z(\phi^s+\phi^t))]\right.\non &&\left.\times\alpha_s(t^\prime_h)S_t(x_3)h(\alpha^\prime_a,\beta^\prime_h,b,b_3)\exp[-S_{gh}(t^\prime_h)]\right\},\\
\mathcal{F}^{\prime LR}_{a}&=&-\mathcal{F}^{\prime LL}_{a}, \\
\mathcal{F}^{\prime SP}_{a}&=&-16\pi C_Ff_B m_{B}^4/\sqrt{2N_C}\int_0^1 d z  dx_{3} \int_{0}^{\infty} b db b_3 db_3\non && 
\times\left\{\left[2\sqrt\eta(\eta-1)\phi^s\phi^A_\pi-r_\pi \phi^0((\eta-1)x_3+1)(\phi^P+\phi^T)-\eta r_\pi \phi^0(\phi^P-\phi^T)\right]\right.\non && \left.\times \alpha_s(t^\prime_{g})S_t(z)h(\alpha^\prime_a,\beta^\prime_g,b_3,b)\exp[-S_{gh}(t^\prime_g)]\right.\non &&\left.-\left[\sqrt\eta z(1-\eta)\phi^A_\pi(\phi^s-\phi^t)+2r_\pi(\eta(z-1)+1)\phi^0\phi^P_\pi \right]\right.\non && \left.\times\alpha_s(t^\prime_h)S_t(x_3)h(\alpha^\prime_a,\beta^\prime_h,b,b_3)\exp[-S_{gh}(t^\prime_h)]\right\},
\en
where the hard scales $t^\prime_i$ with $i=a,\cdots,h$ and the Sudakov factors can be obtained from Eqs. (\ref{scale})-(\ref{suda}) respectively, with the following replacements
\begin{equation}
\begin{aligned}
& \beta_a\to \beta^\prime_a=r_b^2-1+(1-\eta)x_3, && \beta_b\to\beta^\prime_b=x_B(1-\eta),  \\
& \beta_c\to\beta^\prime_c=(x_B+z-1)(\eta+(1-\eta)x_3), &&\beta_d\to\beta^\prime_d=(x_B-z)(1-\eta)x_3, \\
&\beta_e\to\beta^\prime_e=r_b^2+(1-z)(1-x_B-(1-\eta)(1-x_3)),&& \beta_f\to\beta^\prime_f=z(x_B-(1-\eta)(1-x_3)),  \\
& \beta_g\to\beta^\prime_g=(1-\eta)x_3-1, && \beta_h\to\beta^\prime_h=-z(1-\eta),\\
&\alpha_e\to\alpha^\prime_e=x_B(1-\eta)x_3 , \;\;\;\;\; b\leftrightarrow b_3, &&\alpha_a\to\alpha^\prime_a=-z(1-\eta)(1-x_3).
\end{aligned}
\end{equation}
\section{Total amplitudes}
Combining the amplitudes from the different Feynman diagrams, one can obtained the total decays amplitudes for our considered decays, which are listed as follows
\begin{small}
\be
A(B^{+}\rightarrow a_0^{+}(\to h_1h_2)\pi^0) &=& \frac{G_F}{2} \Bigg\{
V_{ub}^{*}V_{ud} \left[ \mathcal{F}^{LL}_e a_2 + \mathcal{M}^{LL}_e C_2 + \mathcal{M}^{LL}_a C_1 + (\mathcal{F}^{LL}_a + \mathcal{F}^{\prime LL}_e) a_1\right. \nonumber \\
&& \left. + (\mathcal{M}^{\prime LL}_e + \mathcal{M}^{\prime LL}_a) C_1 + \mathcal{F}^{\prime LL}_a a_1 \right] - V_{tb}^*V_{td} \left[ \mathcal{F}^{LL}_e (-a_4+ \frac{5}{3}C_9\right. \nonumber \\
&& \left. + C_{10}) + \mathcal{F}^{LR}_e (\frac{3}{2}C_7 + \frac{1}{2}C_8) + \mathcal{F}^{SP}_e (-a_6 + \frac{1}{2}a_8) + \mathcal{M}^{LL}_e (-C_3\right. \nonumber \\
&& \left. + \frac{3}{2}a_{10})  + \mathcal{M}^{LR}_e (-C_5 + \frac{1}{2}C_7) + \mathcal{M}^{SP}_e \frac{3}{2}C_8 + \mathcal{M}^{LL}_a (C_3 + C_9)\right. \nonumber \\
&& \left.  + \mathcal{M}^{LR}_a (C_5 + C_7) + \mathcal{F}^{LL}_a (a_4 + a_{10}) + \mathcal{F}^{SP}_a (a_6 + a_8) 
 + \mathcal{F}^{\prime LL}_e \right.\nonumber \\&& \left.\times
(a_4 + a_{10}) + \mathcal{F}^{\prime SP}_e (a_6 + a_8) + \mathcal{M}^{\prime LL}_e (C_3 + C_9) + \mathcal{M}^{\prime LR}_e \right. \nonumber \\&& \left.
\times(C_5 + C_7) + \mathcal{M}^{\prime LL}_a (C_3 + C_9)  + \mathcal{M}^{\prime LR}_a (C_5 + C_7) \right. \nonumber \\
 && \left. + \mathcal{F}^{\prime LL}_a (a_4 + a_{10})+ \mathcal{F}^{\prime SP}_a (a_6+a_8) \right]
\Bigg\},\\
A(B^{0}\rightarrow a_0^{-}(\to h_1h_2)\pi^+)&=&\frac{G_F}{\sqrt{2}}\left\{ 
V_{ub}^{*}V_{ud} \left[\mathcal{F}^{LL}_e a_1 + \mathcal{M}^{LL}_e C_1 + \mathcal{M}^{\prime LL}_a C_2 + \mathcal{F}^{\prime LL}_{a} a_2 \right] \right. \nonumber \\
&& \left. - V_{tb}^{*}V_{td} \left[\mathcal{F}^{LL}_e (a_4 + a_{10}) + \mathcal{F}^{SP}_e (a_6 + a_8) + \mathcal{M}^{LL}_e (C_3 + C_9)\right. \right. \nonumber \\
&& \left. \left. + \mathcal{M}^{LR}_e (C_5 + C_7) + \mathcal{M}^{LL}_a (C_3 - \frac{1}{2} C_9 + C_4 - \frac{1}{2} C_{10}) \right. \right. \nonumber \\&& \left. \left.
+ \mathcal{M}^{LR}_a (C_5 -\frac{1}{2}C_7) + \mathcal{M}^{SP}_a(C_6 - \frac{1}{2}C_8) + \mathcal{F}^{LL}_a (\frac{4}{3} C_3 + \frac{4}{3} C_4 \right. \right. \nonumber \\
&& \left. \left.
-\frac{2}{3}C_9 - \frac{2}{3} C_{10}) + \mathcal{F}^{LR}_a (a_5-\frac{1}{2}a_7) + \mathcal{F}^{SP}_a (a_6- \frac{1}{2} a_8 ) \right. \right. \nonumber \\
&& \left. \left.
+\mathcal{M}^{\prime LL}_a (C_4 + C_{10})  +\mathcal{M}^{\prime SP}_a (C_6 + C_8)+\mathcal{F}^{\prime LL}_{a} (a_3+ a_9)\right. \right. \nonumber \\
&& \left. \left.
+ \mathcal{F}^{\prime SP}_{a} (a_5+ a_7) \right] \right\},\\
A(B^{0} \rightarrow a_0^{0} (\to h_1h_2)\pi^0) &=& \frac{G_F}{2\sqrt{2}} \left\{V_{ub}^*V_{ud} \left[ \mathcal{F}^{LL}_e a_2 + \mathcal{M}^{LL}_e C_2 + \mathcal{M}^{LL}_a C_2 + \mathcal{F}^{LL}_a a_2 \right. \right.\nonumber \\
&& \left. \left. + \mathcal{F}^{\prime LL}_e a_2 + \mathcal{M}^{\prime LL}_e C_2 + \mathcal{M}^{\prime LL}_a C_2 + \mathcal{F}^{\prime LL}_a a_2 \right]- V_{tb}^* V_{td} \right.\nonumber \\
&& \left. 
\times \left[ \mathcal{F}^{LL}_e( \frac{5}{3}C_9 + C_{10}-a_4) + \mathcal{F}^{LR}_e (\frac{3}{2}C_7 + \frac{1}{2}C_8)  \right.\right.\nonumber \\	&&  \left.\left.+ \mathcal{F}^{SP}_e (\frac{1}{2}a_8-a_6)+ \mathcal{M}^{LL}_e (\frac{3}{2}a_{10}-C_3)+ \mathcal{M}^{LR}_e (\frac{1}{2}C_7-C_5) \right. \right.\nonumber \\
&&\left. \left.  + \mathcal{M}^{SP}_e \frac{3}{2}C_8+ \mathcal{M}^{LL}_a (\frac{3}{2}a_{10}-C_3)  + \mathcal{M}^{LR}_a (\frac{1}{2}C_7-C_5) \right. \right.\nonumber \\
&&\left. \left.
+ \mathcal{M}^{SP}_a \frac{3}{2}C_8 + \mathcal{F}^{LL}_a (\frac{5}{3}C_9 + C_{10}-a_4)  + \mathcal{F}^{LR}_a \frac{3}{2}a_7 \right. \right.\nonumber \\
&& \left. \left.
+ \mathcal{F}^{SP}_a (\frac{1}{2}a_8-a_6) + \mathcal{F}^{\prime LL}_e (\frac{5}{3}C_9 + C_{10}-a_4 ) + \mathcal{F}^{\prime LR}_e \frac{3}{2}a_7 \right. \right.\nonumber \\
&& \left. \left. + \mathcal{F}^{\prime SP}_e (\frac{1}{2}a_8-a_6) + \mathcal{M}^{\prime LL}_e (\frac{3}{2}a_{10}-C_3) \right. \right.\nonumber \\
&& \left. \left. + \mathcal{M}^{\prime LR}_e(\frac{1}{2}C_7-C_5) 
+\mathcal{M}^{\prime SP}_e \frac{3}{2}C_8+\mathcal{M}^{\prime LL}_a ( \frac{3}{2}a_{10}-C_3) \right. \right.\nonumber \\&&\left. \left.
+\mathcal{M}^{\prime LR}_a (\frac{1}{2}C_7-C_5)  
+ \mathcal{M}^{\prime SP}_a \frac{3}{2}C_8 + \mathcal{F}^{\prime LL}_a (\frac{5}{3}C_9 + C_{10}-a_4)  
\right. \right.\nonumber \\&& \left. \left. + \mathcal{F}^{\prime LR}_a (\frac{3}{2}a_7)+ \mathcal{F}^{\prime SP}_a (-a_6 + \frac{1}{2}a_8) \right] \right\},\\
A(B^{0}\rightarrow a_0^{+}(\to h_1h_2)\pi^-) &=& \frac{G_F}{\sqrt{2}} \left\{
V_{ub}^{*}V_{ud} \left[\mathcal{F}^{LL}_e a_1 + \mathcal{M}^{LL}_e C_1 + \mathcal{M}^{\prime LL}_a C_2 + \mathcal{F}^{\prime LL}_a a_2\right] \right. \nonumber \\
&& \left. - V_{tb}^*V_{td} \left[ \mathcal{F}^{LL}_e (a_4+ a_{10}) + \mathcal{F}^{SP}_e(a_6 + a_8 ) + \mathcal{M}^{LL}_e (C_3 + C_9)  \right. \right. \nonumber \\&& \left. \left.
+ \mathcal{M}^{LR}_e (C_5 + C_7) + \mathcal{M}^{LL}_a (C_3 - \frac{1}{2} C_9 + C_4 - \frac{1}{2} C_{10})\right.\right. \nonumber \\&&\left.\left.
+ \mathcal{M}^{LR}_a (C_5 - \frac{1}{2}C_7) +\mathcal{M}^{SP}_a(C_6-\frac{1}{2}C_8) + \mathcal{F}^{LL}_a ( \frac{4}{3} C_3  \right. \right. \nonumber \\
&& \left. \left.
+ \frac{4}{3} C_4-\frac{2}{3} C_9 - \frac{2}{3} C_{10}) + \mathcal{F}^{LR}_a (a_5 - \frac{1}{2}a_7) + \mathcal{F}^{SP}_a (a_6- \frac{1}{2} a_8 )\right. \right. \nonumber \\&& \left. \left.
+\mathcal{M}^{\prime LL}_a(C_4+C_{10}) +\mathcal{M}^{\prime SP}_a(C_6+C_8) +\mathcal{F}^{\prime LL}_a(a_3 +a_9)
\right. \right. \nonumber \\&& \left. \left.
+\mathcal{F}^{\prime LR}_a(a_5 +a_7)\right] \right\}, \\
A(B^{+}\rightarrow a_0^0(\to h_1h_2)\pi^{+})&=& \frac{G_F}{2}\left\{V_{ub}^{*}V_{ud} \left[F_e^{LL} a_1+M _e^{LL} C_1+ M _a^{LL} C_1+ F_a^{L L} a_1+ F_e^{\prime L L} a_2\right.\right.\nonumber \\
&& \left.+ M _e^{L L} C_2+ M _a^{\prime L L} C_1+ F _a^{\prime L L} a_1\right]-V_{t b}^* V_{t d}\left[ F _e^{L L}\left(a_4+a_{10}\right)\right.\nonumber \\
&& + F _e^{S P}\left(a_6+a_8\right)+ M _e^{L L}\left(C_3+\frac{1}{2} C_9\right)+ M _e^{L R}\left(C_5+C_7\right)\nonumber \\
&& + M _a^{L L}\left(C_3+C_9\right)+ M _a^{L R}\left(C_5+C_7\right)+ F _a^{L L}\left(a_4+a_{10}\right)\nonumber \\
&& + F _a^{SP}\left(a_6+a_8\right)+ F_e^{\prime L L}\left(-a_4+\frac{5}{3} C_9+C_{10}\right)+ F_e^{\prime LR}\frac{3}{2} a_7\nonumber\\&& 
+F_e^{\prime SP}\left(\frac{1}{2} a_8-a_6\right)+ M_e^{\prime LL}\left(-C_3+\frac{3}{2} a_{10}\right)+ M_e^{\prime LR}\left(\frac{1}{2} C_7-C_5\right)\nonumber\\
&&+M _e^{\prime S P}\frac{3}{2} C_8+M_a^{\prime LR}\left(C_5+C_7\right)+ F_a^{\prime LL}\left(a_4+a_{10}\right)\nonumber \\&& \left.\left.
+M_a^{\prime LL}\left(C_3+C_9\right)+ F_a^{\prime SP}\left(a_6+a_8\right)\right]\right\},\\
A(B^{0}\rightarrow a_0^{0}(\to h_1h_2)K^0)&=&\frac{G_F}{2}\left\{ 
V_{ub}^*V_{us} \left[
\mathcal{F}^{\prime LL}_e a_2+
\mathcal{M}^{\prime LL}_e C_2
\right] - V_{tb}^*V_{ts} \left[
\mathcal{F}^{\prime LL}_e \frac{3}{2}a_9\right. \right.\nonumber\\
&& \left.\left.
+\mathcal{F}^{\prime LR}_e \frac{3}{2}a_7+
\mathcal{M}^{\prime LL}_e \frac{3}{2}C_{10} +
\mathcal{M}^{\prime SP}_e \frac{3}{2}C_8
 + 
\mathcal{F}^{LL}_e(a_4-\frac{1}{2}a_{10}) \right.\right.\nonumber\\
&& \left.\left.
+\mathcal{F}^{SP}_e (a_6-a_8)+
\mathcal{M}^{LL}_e (C_3-\frac{1}{2}C_9)+\mathcal{M}^{LR}_e (C_5-\frac{1}{2}C_7) \right. \right.\nonumber\\
&& \left.\left.
+\mathcal{M}^{LL}_a (C_3-\frac{1}{2}C_9)+
\mathcal{M}^{LR}_a (C_5-\frac{1}{2}C_7) 
\right.\right.\nonumber\\&& \left.\left.+
\mathcal{F}^{LL}_a (a_4-\frac{1}{2}a_{10})+\mathcal{F}^{SP}_a (a_6 - \frac{1}{2}a_8)\right]\right\},\\
A(B^{+}\rightarrow a_0^{0}(\to h_1h_2)K^+)&=&\frac{G_F}{2}\left\{ 
V_{ub}^{*}V_{us} \left[
\mathcal{F}^{LL}_e a_1 +\mathcal{M}^{LL}_e C_1 +\mathcal{M}^{LL}_a C_1 +
\mathcal{F}^{LL}_a a_1 \right.\right.\non&&\left.\left.
+\mathcal{F}^{\prime LL}_e a_2 +\mathcal{M}^{\prime LL}_e C_2\right] - V_{tb}^{*}V_{ts} \left[
\mathcal{F}^{LL}_e (a_4 + a_{10})\right.\right.\non&&\left.\left.
+\mathcal{F}^{SP}_e (a_6+a_8) +
\mathcal{M}^{LL}_e (C_3 + C_9) +
\mathcal{M}^{LR}_e (C_5 + C_7)\right.\right.\non&&\left.\left.
+\mathcal{M}^{LL}_a (C_3 + C_9) +
\mathcal{M}^{LR}_a (C_5 + C_7) +
\mathcal{F}^{LL}_a(a_4+a_{10})\right.\right.\non&&\left.\left.
+\mathcal{F}^{SP}_a (a_6+a_8) +
\mathcal{F}^{\prime LL}_e \frac{3}{2}a_9 +
\mathcal{F}^{\prime LR}_e \frac{3}{2}a_8\right.\right.\non&&\left.\left.
+\mathcal{M}^{\prime LL}_e \frac{3}{2}C_{10} +
\mathcal{M}^{\prime SP}_e \frac{3}{2}C_8\right]\right\},\\
A(B^{+}\rightarrow a_0^{+}(\to h_1h_2) K^0) &=& \frac{G_F}{\sqrt{2}} \left\{ 
V^*_{ub}V_{us} \left[\mathcal{M}^{LL}_a C_1 + \mathcal{F}^{LL}_a a_1 \right] - V^*_{tb}V_{ts} \left[ \mathcal{F}^{LL}_e (a_4 - \frac{1}{2}a_{10})\right.\right. \non 
&& \left.\left.  + \mathcal{F}^{SP}_e (a_6 - \frac{1}{2}a_8)+ \mathcal{M}^{LL}_e (C_3 
-\frac{1}{2}C_9)+ \mathcal{M}^{LR}_e (C_5 - \frac{1}{2}C_7 ) \right.\right.\non 
&&\left.\left.+ \mathcal{M}^{LL}_a (C_3+C_9) + \mathcal{M}^{LR}_a (C_5
+C_7)+ \mathcal{F}^{LL}_a (a_4 + a_{10}).\right.\right.\non 
&&\left.\left. + \mathcal{F}^{SP}_a (a_6+a_8) \right] \right\}, \\
A(B^{0}\rightarrow a_0^{-}(\to h_1h_2)K^+)&=&\frac{G_F}{\sqrt{2}} \left\{ 
V^*_{ub}V_{us} \left[ \mathcal{F}^{LL}_e a_1 + \mathcal{M}^{LL}_e C_1 \right]-V^*_{tb}V_{ts} \left[\mathcal{F}^{LL}_e (a_4+a_{10}) \right.\right. \non
&& \left.\left. +\mathcal{F}^{SP}_e (a_6+a_8) +\mathcal{M}^{LL}_e(C_3+C_9) + \mathcal{M}^{LR}_e ( C_5 + C_7 ) \right. \right. \non
&& \left. \left.+ \mathcal{M}^{LL}_a ( C_3 - \frac{1}{2}C_9 ) + \mathcal{M}^{LR}_a ( C_5 - \frac{1}{2}C_7 ) \right. \right. \non
&& \left. \left. + \mathcal{F}^{LL}_a (a_4 - \frac{1}{2}a_{10} ) + \mathcal{F}^{SP}_a (a_6 - \frac{1}{2}a_8) \right] \right\},
\en
\end{small}
where $h_1h_2$ refers to $\pi\eta(K\bar K)$ pair, $G_F$ is the Fermi constant and the effective coefficients $a_i$ are defined in terms of the Wilson coefficients as~\cite{Ali}
\begin{align}
a_1 &=C_2+C_1/3,\; a_3=C_3+C_4/3,\;  a_5=C_5+C_6/3,\;  a_7=C_7+C_8/3,\;  a_9=C_9+C_{10}/3, \notag \\[0.15cm]
a_2 &=C_1+C_2/3,\; a_4=C_4+C_3/3,\;  a_6=C_6+C_5/3,\;  a_8=C_8+C_7/3,\;  a_{10}=C_{10}+C_9/3.
\end{align}



\begin{thebibliography}{99}
\bibitem{Hanhart}
C.~Hanhart, B.~Kubis and J.~R.~Pelaez,
Phys. Rev. D \textbf{76}, 074028 (2007)
[arXiv:0707.0262 [hep-ph]].
\bibitem{Sekihara}
T.~Sekihara and S.~Kumano,
Phys. Rev. D \textbf{92}, 034010 (2015)
[arXiv:1409.2213 [hep-ph]].
\bibitem{wwang}
W.~Wang,
Phys. Lett. B \textbf{759}, 501 (2016)
[arXiv:1602.05288 [hep-ph]].
\bibitem{close}
F.~E.~Close and A.~Kirk,
Phys. Lett. B \textbf{489}, 24-28 (2000)
[arXiv:hep-ph/0008066].
\bibitem{achasov1}
N.~N.~Achasov and G.~N.~Shestakov,
Phys. Rev. Lett. \textbf{92}, 182001 (2004)
[arXiv:hep-ph/0312214].
\bibitem{wujj}
J.~J.~Wu, Q.~Zhao and B.~S.~Zou,
Phys. Rev. D \textbf{75}, 114012 (2007)
[arXiv:0704.3652 [hep-ph]].
\bibitem{jaffe}
R.~L.~Jaffe,
Phys. Rev. D \textbf{15}, 267 (1977).
\bibitem{amsler}
C.~Amsler \textit{et al.} [Crystal Barrel],
Phys. Lett. B \textbf{333}, 277-282 (1994).
\bibitem{babar}
B.~Aubert \textit{et al.} [BaBar],
Phys. Rev. D \textbf{70}, 111102 (2004)
[arXiv:hep-ex/0407013 [hep-ex]].
\bibitem{babar2}
B.~Aubert \textit{et al.} [BaBar],
Phys. Rev. D \textbf{75}, 111102 (2007) [arXiv:hep-ex/0703038 [hep-ex]].
\bibitem{LHCb1}
R.~Aaij \textit{et al.} [LHCb],
Phys. Rev. D \textbf{88}, 072005 (2013)
[arXiv:1308.5916 [hep-ex]].
\bibitem{bes3}
M.~Ablikim \textit{et al.} [BESIII],
Phys. Rev. Lett. \textbf{121}, 081802 (2018)
[arXiv:1803.02166 [hep-ex]].
\bibitem{cleo}
P.~Rubin \textit{et al.} [CLEO],
Phys. Rev. D \textbf{78}, 072003 (2008)
[arXiv:0807.4545 [hep-ex]].
\bibitem{LHCb2}
R.~Aaij \textit{et al.} [LHCb],
Phys. Rev. D \textbf{93}, 052018 (2016)
[arXiv:1509.06628 [hep-ex]].
\bibitem{bes31}
M.~Ablikim \textit{et al.} [BESIII],
[arXiv:2404.09219 [hep-ex]].
\bibitem{frxu}
H.~Y.~Cheng, C.~W.~Chiang and F. R. Xu,
[arXiv:2408.13942 [hep-ph]].
\bibitem{achasov}
N.~N.~Achasov and G.~N.~Shestakov,
Phys. Rev. D \textbf{110}, 016025 (2024)
[arXiv:2405.13332 [hep-ph]].
\bibitem{chenghy1}
H.~Y.~Cheng, C.~K.~Chua and K.~C.~Yang,
Phys. Rev. D \textbf{73}, 014017 (2006)
[arXiv:hep-ph/0508104 [hep-ph]].
\bibitem{chenghy2}
H.~Y.~Cheng, C.~K.~Chua, K.~C.~Yang and Z.~Q.~Zhang,
Phys. Rev. D \textbf{87}, 114001 (2013)
[arXiv:1303.4403 [hep-ph]].
\bibitem{chenghy3}
H.~Y.~Cheng, C.~K.~Chua and K.~C.~Yang,
Phys. Rev. D \textbf{77}, 014034 (2008)
[arXiv:0705.3079 [hep-ph]].
\bibitem{shenyl}
Y.~L.~Shen, W.~Wang, J.~Zhu and C.~D.~Lu,
Eur. Phys. J. C \textbf{50}, 877 (2007) [arXiv:hep-ph/0610380 [hep-ph]].
\bibitem{zhangzq1}
Z.~Q.~Zhang and Z.~J.~Xiao,
Chin. Phys. C \textbf{34}, 528 (2010) [arXiv:0904.3375 [hep-ph]].
\bibitem{zhangzq2}
Z.~Q.~Zhang, Phys. Rev. D \textbf{83}, 054001 (2011)
[arXiv:1106.0368 [hep-ph]].
\bibitem{zouzt}
Z.~T.~Zou, Y.~Li and X.~Liu,
Eur. Phys. J. C \textbf{77}, 870 (2017)
[arXiv:1704.03967 [hep-ph]].
\bibitem{chaij}
J.~Chai, S.~Cheng and A.~J.~Ma,
Phys. Rev. D \textbf{105}, 033003 (2022)
[arXiv:2109.00664 [hep-ph]].
\bibitem{zhour}
Z.~Rui, Y.~Q.~Li and J.~Zhang,
Phys. Rev. D \textbf{99}, 093007 (2019)
[arXiv:1811.12738 [hep-ph]].
\bibitem{H. N. Li}
H.~n.~Li,
Prog. Part. Nucl. Phys. \textbf{51}, 85-171 (2003)
[arXiv:hep-ph/0303116 [hep-ph]].
\bibitem{T. Kurimoto}
T.~Kurimoto, H.~n.~Li and A.~I.~Sanda,
Phys. Rev. D \textbf{65}, 014007 (2002)
[arXiv:hep-ph/0105003 [hep-ph]].
\bibitem{WFWangHNLiWWangandCDLu}
W.~F.~Wang, H.~n.~Li, W.~Wang and C.~D.~L\"u,
Phys. Rev. D \textbf{91}, 094024 (2015)
[arXiv:1502.05483 [hep-ph]].
\bibitem{CHChenHNLi}
C.~H.~Chen and H.~n.~Li,
Phys. Lett. B \textbf{561}, 258-265 (2003)
[arXiv:hep-ph/0209043 [hep-ph]].
\bibitem{ZRuiWFWang}
Z.~Rui and W.~F.~Wang,
Phys. Rev. D \textbf{97}, 033006 (2018)
[arXiv:1711.08959 [hep-ph]].
\bibitem{UMeinerWWang}
U.~G.~Mei\ss{}ner and W.~Wang,
Phys. Lett. B \textbf{730}, 336-341 (2014)
[arXiv:1312.3087 [hep-ph]].
\bibitem{Aaij}
R.~Aaij \textit{et al.} [LHCb],
Phys. Rev. D \textbf{88}, 072005 (2013)
[arXiv:1308.5916 [hep-ex]].
\bibitem{Abele}
A.~Abele, S.~Bischoff, P.~Blum, N.~Djaoshvili, D.~Engelhardt, A.~Herbstrith, C.~Holtzhaussen, M.~Tischhauser, J.~Adomeit and B.~Kammle, \textit{et al.}
Phys. Rev. D \textbf{57}, 3860 (1998).
\bibitem{Aaij2}
R.~Aaij \textit{et al.} [LHCb],
Phys. Rev. D \textbf{93}, 052018 (2016)
[arXiv:1509.06628 [hep-ex]].
\bibitem{zhaoyc}
Y.~C.~Zhao, Z.~Q.~Zhang, Z.~Y.~Zhang, Z.~J.~Sun and Q.~B.~Meng,
Chin. Phys. C \textbf{47}, 073104 (2023)
[arXiv:2304.13286 [hep-ph]].
\bibitem{pdg}
R.~L.~Workman \textit{et al.} [Particle Data Group],
PTEP \textbf{2022}, 083C01 (2022).
\bibitem{zhenghaiyang}
H.~Y.~Cheng, C.~K.~Chua, K.~C.~Yang and Z.~Q.~Zhang,
Phys. Rev. D \textbf{87}, 114001 (2013)
[arXiv:1303.4403 [hep-ph]].
\bibitem{zhangzhiqing}
Z.~Q.~Zhang and Z.~J.~Xiao,
Chin. Phys. C \textbf{34}, 528 (2010)
[arXiv:0904.3375 [hep-ph]].
\bibitem{Ali}
A.~Ali, G.~Kramer and C.~D.~Lu,
Phys. Rev. D \textbf{58}, 094009 (1998)
[arXiv:hep-ph/9804363 [hep-ph]].
\end{thebibliography}

\end{document}